\documentclass[reprint,amsmath,amssymb, prb, aps, showpacs]{revtex4-1}

\usepackage{amssymb,amsmath,color,amsthm,longtable,graphicx}
\usepackage{hyperref}
\usepackage{tikz}
\usetikzlibrary{arrows,shapes}

\newcommand{\Integer}{\mathbb{Z}}
\newcommand{\Complex}{\mathbb{C}}
\newcommand{\Real}{\mathbb{R}}
\newcommand{\cZ}{\mathcal{Z}}
\newcommand{\g}{\mathfrak{g}}
\newcommand{\cB}{\mathcal{B}}
\newcommand{\cH}{\mathcal{H}}  
\DeclareMathOperator{\End}{\text{End}}
\newcommand{\cG}{\mathcal{G}}  
\newcommand{\idop}{1\!\!1}
\newcommand{\cK}{\mathcal{K}}
\DeclareMathOperator{\tr}{\text{tr}}
\DeclareMathOperator{\Ad}{{\text{Ad}}}
\DeclareMathOperator{\Hom}{\text{Hom}}
\newtheorem{proposition}{Proposition}

\begin{document}

\title{On topological phases of spin chains}

\author{Kasper Duivenvoorden}
 \email{Kasper@thp.uni-koeln.de}
\author{Thomas Quella}
 \email{Thomas.Quella@uni-koeln.de}
 \affiliation{Institute of Theoretical Physics, University of Cologne\\
 Z\"ulpicher Stra\ss{}e 77, D-50937 Cologne, Germany 
}

\date{\today}

\begin{abstract}
  Symmetry protected topological phases of one-dimensional spin
  systems have been classified using group cohomology. In this
  paper, we revisit this problem for general spin chains which are
  invariant under a continuous on-site symmetry group $G$. We evaluate
  the relevant cohomology groups and find that the topological phases
  are in one-to-one correspondence with the elements of the
  fundamental group of $G$ if $G$ is compact, simple and
  connected and if no additional symmetries are imposed.
  For spin chains with symmetry $PSU(N)=SU(N)/\Integer_N$
  our analysis implies the existence of $N$ distinct topological
  phases. For symmetry groups of orthogonal, symplectic or
  exceptional type we find up to four different phases. Our work
  suggests a natural generalization of Haldane's conjecture beyond
  $SU(2)$.
\end{abstract}

\pacs{03.65.Vf, 75.10.Pq, 03.65.Fd}

\maketitle


\parskip2pt

\section{Introduction}

  The integer quantum Hall effect is the best known example of a
  condensed matter system where a physical observable -- the electric
  conductance -- can be expressed in terms of a discrete,
  $\Integer$-valued topological
  invariant. The interest in such topological phases of matter was
  renewed with the prediction of a spin quantum Hall effect and an
  associated $\Integer_2$ topological invariant in graphene with
  time-reversal invariant spin orbit interactions.
  \cite{Kane:PhysRevLett.95.146802} Soon after, a generalization of
  the spin quantum Hall effect to three dimensions was found.
  \cite{Fu:PhysRevLett.98.106803} By now, a comprehensive
  classification of non-interacting fermionic systems is available
  which describes various types of topological insulators and
  superconductors.
  \cite{Schnyder:2008,Kitaev:2009mg,Ryu:1367-2630-12-6-065010,LeClair:2012arXiv1205.3810L}
  These results have been motivated by the symmetry
  classification of quadratic random Hamiltonians \`{a} la Altland and
  Zirnbauer. \cite{Altland:1997zz,Heinzner:2004xj}

  More recently, the focus shifted towards interacting systems. Due to
  strong correlations between the electrons, the notion of a band
  structure ceases to be valid and alternative methods to detect and
  to classify topological phases have to be sought. The bulk-boundary
  correspondence, i.e.\ the prediction of massless surface modes at
  the interface between two topologically distinct bulk systems,
  serves as a useful guiding principle. Evidence may also be gained
  from characteristic entanglement spectra
  \cite{Li:PhysRevLett.101.010504,Fidkowski:PhysRevLett.104.130502}
  which contain information about potential surface modes by
  introducing virtual interfaces into the system or from
  single-particle Green's functions.
  \cite{Manmana:2012arXiv1205.5095M}
  The first systematic studies of topological phases of interacting
  fermions have been concerned with Majorana chains.
  \cite{Fidkowski:PhysRevB.81.134509,Fidkowski:2011PhRvB..83g5103F,Turner:2011PhRvB..83g5102T}
  For these chains it was shown that the $\Integer$-classification of
  the corresponding non-interacting symmetry class is reduced to a
  $\Integer_8$-classification. Similar results for other systems have
  been obtained in Refs.\
  \onlinecite{Ryu:2012arXiv1202.5805Y,Ryu:2012PhRvB..85x5132R}.

  Topologically non-trivial phases are not confined to fermionic
  systems but they also arise naturally in bosonic models, e.g.\ in
  interacting spin systems. A specific deformation of the
  $SU(2)$-invariant antiferromagnetic Heisenberg spin chain with spin
  $S=1$, the so-called AKLT spin chain,
  \cite{Affleck:PhysRevLett.59.799,Affleck:1987cy} was probably the
  first example of this type. This system
  exhibits the following hallmarks of a topological phase: with
  periodic boundary conditions there is a gap above a unique ground
  state, \cite{Affleck:1987cy} one has a bulk-boundary correspondence:
  open boundary conditions imply massless edge modes carrying a
  topological quantum number, \cite{Hagiwara:PhysRevLett.65.3181} the
  ground state leads to a characteristic entanglement spectrum
  \cite{Katsura:2008JPhA...41m5304K,Pollmann:PhysRevB.81.064439} and
  last but not least there exists a non-local string order parameter.
  \cite{DenNijs:PhysRevB.40.4709}

  Various extensions of the AKLT setup to higher rank groups
  and supersymmetric systems have been considered, see e.g.\ Refs.\ 
  \onlinecite{Affleck:1987cy,Greiter:PhysRevB.75.184441,Schuricht:PhysRevB.78.014430,Tu:PhysRevB.78.094404,Arovas:PhysRevB.79.224404}.
  Other generalizations include $q$-deformations of the symmetry group
  which can be used to describe anisotropic spin chains.
  \cite{Klumper:1992ZPhyB..87..281K,Batchelor:1994IJMPB...8.3645B,Fannes:10.1007/BF02101525}
  In all these examples the matrix product (or valence bond) state
  formalism plays a crucial role.
  \cite{Fannes:1989ns,Fannes:1990px,Fannes:1990ur,Perez-Garcia:2007:MPS:2011832.2011833}
  Indeed, the latter  is extremely useful when classifying symmetry
  protected topological phases of one-dimensional spin systems since
  boundary and entanglement properties are almost trivial to access.
  \cite{Chen:PhysRevB.83.035107,Schuch:1010.3732v3,Chen:PhysRevB.84.235128}
  In the meantime, also proposals have been presented how to address
  fermionic systems in this framework and how to lift the
  classification to higher dimensional systems using projective
  entangled pairs and, more generally, tensor network states 
  \cite{Schuch:1010.3732v3,Chen:2011arXiv1106.4772C,Gu:2012arXiv1201.2648G}
  (see also Ref.\ \onlinecite{Bachmann:2012CMaPh.309..835B} for a
  $C^\ast$-algebraic point of view).

  In the present paper we are considering gapped antiferromagnetic
  spin chains which are invariant under the action of an arbitrary
  compact connected simply-connected simple Lie group $G$. In
  contrast, we do not impose any additional symmetries such as
  time-reversal or inversion symmetry. Under these conditions, the
  general classification predicts that the distinct topological phases
  are labeled by the elements of a certain cohomology group.
  \cite{Chen:PhysRevB.83.035107,Schuch:1010.3732v3}
  Depending on the concrete system under study, the relevant
  cohomology groups are $H^2\bigl(G/\Gamma,U(1)\bigr)$ where
  $\Gamma\subset\cZ(G)$ denotes a central subgroup of $G$. Elements of
  this cohomology label the distinct classes of projective
  representations of $G/\Gamma$. The group $\Gamma$ is determined by
  the representations of $G$ which are used to describe the physical
  spins.

  To our knowledge, so far explicit results on the cohomology
  groups $H^2\bigl(G/\Gamma,U(1)\bigr)$ have only appeared in the
  condensed matter literature for the orthogonal groups
  $SO(N)=Spin(N)/\Integer_2$ where two topological phases have been
  found.\cite{Haegeman:1201.4174v1} In addition, the cohomologies for
  the classical groups $SU(N)$ and $SP(N)$ (corresponding to
  $\Gamma=\{1\}$) have been written down in Ref.\
  \onlinecite{Chen:2011arXiv1106.4772C}. However, the
  corresponding phases all turn out to be topologically trivial, at
  least in one dimension. In our paper, we will fill this gap and show
  that the cohomology group $H^2\bigl(G/\Gamma,U(1)\bigr)$ is
  isomorphic to $\Gamma$, which can also be interpreted as the
  fundamental group of $G/\Gamma$ (see eq.\
  \eqref{eq:EqualityOfGroups}). Hence there are $|\Gamma|$ distinct
  topological phases. This number becomes maximal for $\Gamma=\cZ(G)$
  in which case the resulting group $PG=G/\cZ(G)$ is called the
  projective group associated with $G$. For $PSU(N)$, for instance,
  our result implies the existence of $N$ distinct topological
  phases.

  Besides stating an abstract classification result, we also
  discuss how each non-trivial topological phase can be engineered
  using matrix product states. For this purpose we state an
  explicit formula which determines the projective class of a
  representation of $G$ if it is interpreted as a projective
  representation of $PG$ (see eq.\ \eqref{eq:ProjClass}). The
  topological phases fall into different hierarchies with regard to
  different choices of central subgroups $\Gamma\subset\cZ(G)$. This
  information is sufficient to determine the projective class
  with respect to any of the quotients $G/\Gamma$. While, from
  a mathematical perspective, we are merely summarizing well-known
  facts, we hope that the explicitness of our presentation will be
  useful to the practitioner.

  Our paper ends with a discussion of physical implications. We first
  reveal a physical interpretation for the hierarchy of
  topological phases. More importantly, the mere existence of such a
  hierarchy suggests a natural generalization of Haldane's
  conjecture \cite{Haldane:1983464,Haldane:PhysRevLett.50.1153} to
  arbitrary symmetry groups. In particular, we conjecture the
  existence of confined spinon phases in spin chains with $SO(2N)$
  symmetry and long range interactions.
  Even though spin chains with higher rank symmetry groups like
  $SU(N)$ or $SO(2N)$ are unlikely to be found in real materials,
  there is a chance that the corresponding Hamiltonians can be
  engineered artificially using ultracold atoms in optical lattices.
  \cite{GarciaRipoll:PhysRevLett.93.250405,2010NatPh...6..289G,Hung:2011PhRvB..84e4406H,Nonne:2012arXiv1210.2072N}
  Also, special points in the moduli space of spin chains and spin
  ladders might exhibit an enhanced symmetry. This for instance
  happens for $SU(2)$ spin chains which are known to possess an
  $SU(3)$ symmetric point for a certain value of the
  couplings.\cite{Affleck:1985wb}

  The article is organized as follows. In Section
  \ref{sc:Prerequisites} we present a number of physical and
  mathematical prerequisites. From a physical perspective this
  includes a precise definition of the setup, a brief review of the
  classification of topological phases in terms of the second
  cohomology of the symmetry group and the general definition of
  matrix product states. The mathematical part is concerned with the
  relation between a Lie algebra $\g$ and its various associated
  compact connected Lie groups, which can all be represented as a
  quotient $G/\Gamma$ of a simply-connected universal covering group
  $G$. We introduce the congruence class $[\lambda]$ of an irreducible
  representation $\lambda$ of $\g$. The value of $[\lambda]$ measures
  whether the representation can be lifted to a linear representation
  of $PG$ or not. We also recall the intimate connection between
  central extensions and covering groups.

  Section \ref{sc:TopPhases} contains the main result of the paper: We
  identify the second cohomology of the groups $G/\Gamma$ with their
  fundamental group $\Gamma$, thereby giving a direct classification
  of topological phases. In a case by case study, we afterwards
  determine the number of topological phases and their characteristics
  for each compact connected simple Lie group. Our presentation
  includes explicit formulas for the congruence class of
  representations which may be used to characterize gapless edge
  modes. In Section~\ref{sc:Physics} we return to the physical
  realization of  topologically non-trivial phases in spin chains. We
  give a physical interpretation for the mathematical hierarchy of
  topological phases in terms of a blocking procedure. Otherwise the
  main focus centers around a generalization of Haldane's conjecture
  to spin chains with arbitrary continuous symmetry. Section
  \ref{sc:ColdAtoms} features an application of our formalism to
  $SU(N)$ spin chains that arise in the context of cold atom
  systems. Our results support the observation of
  Ref.~\onlinecite{Nonne:2012arXiv1210.2072N} that non-trivial
  topological phases should be realizable in such systems. Finally,
  Section \ref{sc:Conclusions} provides a summary and concluding
  remarks. In particular, we briefly sketch the modification of our
  classification when space-time symmetries are enforced.

\section{\label{sc:Prerequisites}Physical and mathematical
  prerequisites}

  The first half of this section is used to define 1D spin systems
  with continuous symmetries and to briefly review the classification
  of topological phases in such systems by means of cohomology
  groups. For later convenience we also recall the characterization of
  non-trivial topological phases in terms of massless edge modes. In
  the second half we present some important facts on Lie algebras and
  Lie groups which are well-known in mathematics but required for a
  self-contained presentation of our results. Our main focus is the
  relation between Lie algebras and Lie groups. We discuss which
  groups can be obtained by exponentiating a given Lie algebra $\g$
  and which representations of $\g$ lift to which of these groups --
  possibly projectively. For this purpose we introduce congruence
  classes of $\g$-representations. Finally, we discuss the relation
  between finite coverings of Lie groups and their central
  extensions.

\subsection{Physical setup}

  We base the definition of 1D spin chains on the following data: A
  simple Lie algebra $\g$ of symmetries, a representation $\cH_k$ of
  $\g$ attached to each of the sites $k$ and a Hamiltonian
  $H\in\End_\g(\cH)$ which acts on the total Hilbert space
  $\cH=\cH_1\otimes\cdots\otimes\cH_L$ of the system and which
  commutes with the action of $\g$. In addition, one might wish to
  impose specific boundary conditions (open, periodic, ...) which are
  compatible with the action of $\g$. For physical reasons, the
  Hamiltonian should be local, i.e.\ one should be able to write it as
  a sum $H=\sum_k H_k$ where each summand $H_k$ only affects a finite
  number of sites. Since the quadratic Casimir is the only second
  order invariant of a simple Lie algebra, every Hamiltonian with
  two-body interactions will be a function of the product
  $\vec{S}_k\cdot\vec{S}_{l}$ of the two ``spin operators'' on the
  sites $k$ and $l$.

  Given this setup, it is important to note that $\g$ alone
  does not (necessarily) determine the full symmetry of the system. In
  particular, there might be discrete symmetries (e.g.\ translations but
  also on-site symmetries) which necessarily need to be described by a
  group. They cannot be captured by the symmetry
  algebra $\g$ but may well be relevant for a characterization
  and/or classification of topological phases. Besides the choice of
  $\g$, also the choice of representations $\cH_k$ will play a crucial
  role in the discussion of discrete symmetries. To give just one
  trivial example, translations by one site only have a chance to be a
  symmetry of the system if all spaces $\cH_k$ are chosen to be
  isomorphic and periodic boundary conditions are imposed.

  More important for the purpose of this paper, when lifting the
  symmetry described in terms of the Lie algebra $\g$ to a group
  symmetry $G$ one might have several choices and not all of them will
  lead to a faithful representation of $G$ on the spaces $\cH_k$. A
  simple example is the $S=1$ representation of $SU(2)$ which cannot
  distinguish the two central elements $\pm1\in SU(2)$ and hence only
  corresponds to a faithful representation of
  $SU(2)/\Integer_2=SO(3)$. In Sections \ref{sc:GlobalProperties} and
  \ref{sc:Lifting} and then in Section \ref{sc:TopPhases} below we
  will discuss additional (and less familiar) examples of this
  type. Being aware of subtle differences like the ones just mentioned
  is the key to the classification of topological phases in the
  presence of continuous symmetries.

\subsection{\label{sc:MPSClass}The classification of topological
  phases}
  
  A complete classification of one-dimensional gapped spin systems has
  been obtained in Ref.\ 
  \onlinecite{Chen:PhysRevB.83.035107,Schuch:1010.3732v3,Chen:PhysRevB.84.235128}.
  We use this and the following section to review these results.
  In case one is only interested in topological phases sharing the
  same on-site symmetry group $G$, the classification is particularly
  simple: Different topological classes are in one-to-one
  correspondence with the cohomology group $H^2\bigl(G,U(1)\bigr)$
  (with trivial action of $G$ on $U(1)$). If, in addition, space-time
  symmetries are taken into account, the classification becomes more
  complicated.\cite{Chen:PhysRevB.84.235128} In this
  paper we wish to keep the presentation simple, thus neglecting
  potential space-time symmetries throughout the main part of the
  text. Necessary modifications arising from the presence of
  space-time symmetries will be briefly discussed in the conclusions.

  Before we proceed let us briefly recall the definition of the
  cohomology group $H^2\bigl(G,U(1)\bigr)$. For this purpose let us
  consider maps $\omega:G\times G\to U(1)$ which are solutions to the
  cocycle equation
\begin{align}
  \label{eq:Cocycle}
  \omega(g_1,g_2)\,\omega(g_1g_2,g_3)
  \ =\ \omega(g_2,g_3)\,\omega(g_1,g_2g_3)\ \ .
\end{align}
  The set of cocycles forms an abelian group $\cG$ under pointwise
  multiplication. Furthermore, there are trivial solutions of the
  cocycle condition which, for $f:G\to U(1)$, have the form
\begin{align}
  \label{eq:Coboundary}
  \omega(g_1,g_2)\ =\ f(g_1g_2)/f(g_1)f(g_2)
\end{align}
  Solutions of this form are called coboundaries and they form a
  subgroup $\cK$ of $\cG$. The cohomology group above is defined as
  the quotient $H^2\bigl(G,U(1)\bigr)=\cG/\cK$. In the cases of
  interest this is a finite abelian group (Prop.\,2.2 of
  \onlinecite{Moore1964:MR0171880}).

  Cocycles arise naturally from projective representations of $G$,
  i.e.\ from maps $D:G\to U(N)$ satisfying
\begin{align}
  \label{eq:ProjRep}
  D(g_1)D(g_2)\ =\ \omega(g_1,g_2)\,D(g_1g_2)\ \ .
\end{align}
  From this point of view, the cocycle condition \eqref{eq:Cocycle} is
  just the associativity condition for the multiplication law
  \eqref{eq:ProjRep} while the identification of coboundaries with the
  trivial cocycle arises from the desire to trivialize the
  transformation $D(g)\to f(g)D(g)$.

\begin{figure}
\includegraphics[]{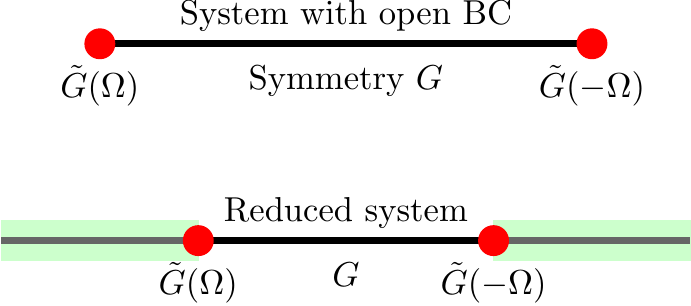}
  \caption{\label{fig:EdgeModes}(Color online) Physical and virtual
    edge modes (red dots) in topologically non-trivial spin
    chains. For simplicity of illustration, the spin chain is depicted
    as a continuous system.}
\end{figure}

  From a physical perspective, the relevance of the second cohomology
  group $H^2\bigl(G,U(1)\bigr)$ can be understood as follows: Each
  element $\Omega\in H^2\bigl(G,U(1)\bigr)$ labels a different central
  extension $\tilde{G}(\Omega)$ of $G$. If $\omega\in\Omega$ is a
  representative of the class $\Omega$, this central extension
  $\tilde{G}(\Omega)$ is defined as the set $G\times U(1)$ with group
  multiplication
\begin{align}
  (g,\alpha)\cdot(h,\beta)
  \ :=\ (gh,\alpha\beta\,\omega(g,h)/\omega(1,1))\ \ .
\end{align}
  One can check that cocycles $\omega$ belonging to the same class
  $\Omega$ give rise to isomorphic central extensions. The choice
  $\omega(g_1,g_2)=1$ corresponds to the trivial central extension
  $\Omega=[0]$. Now the important point is the following: While the
  total system is invariant under the symmetry group $G$, the system
  will exhibit gapless edge modes when considered with open boundary
  conditions.\cite{Chen:PhysRevB.83.035107,Schuch:1010.3732v3}
  The latter transform under one of the enhanced
  symmetries $\tilde{G}(\Omega)$ if the system is in a topologically
  non-trivial phase. If the system has periodic boundary conditions,
  the same reasoning applies. However, now the edge modes are not real
  but they rather appear virtually in the bipartite entanglement
  spectrum after part of the system has been traced out.
  \cite{Katsura:2008JPhA...41m5304K,Pollmann:PhysRevB.81.064439} The
  two possibilities are sketched in Figure \ref{fig:EdgeModes}.

  So far, we have not discussed the class of functions that we wish to
  allow for the cocycles $\omega:G\times G\to U(1)$ and the
  functions $f:G\to U(1)$ entering eqs.\
  \eqref{eq:Cocycle} and \eqref{eq:Coboundary}. For the finite groups
  mostly used in Refs.~\onlinecite{Chen:PhysRevB.83.035107,Schuch:1010.3732v3,Chen:PhysRevB.84.235128}
  there is actually no restriction. However, since our paper is
  concerned with continuous groups one should impose additional
  regularity conditions. Demanding continuity turns out to be too
  restrictive. Indeed, all we need is that linear and projective
  representations are implemented in terms of continuous homomorphisms
  $R:G\to U(N)$ and $D:G\to PU(N)$, respectively, where
  $PU(N)=U(N)/U(1)$. In this formulation, any reference to cocycles is
  missing altogether. In fact, in order to be admissible, the cocycles
  only have to respect a Borel structure on the relevant groups $G$
  and $U(1)$, i.e.\ they have to be measurable functions. Since a
  Borel structure is less restrictive than a topology, this opens the
  possibility for discontinuous jumps on sets of measure
  zero. Fortunately, these rather technical aspects are not relevant
  for the further presentation of the subject. For this reason, we
  refer interested readers to the more detailed expositions available
  in the original literature.\cite{Mackey:1957MR0089999,Mackey:1958MR0098328,Moore1964:MR0171880}

\subsection{\label{sc:MPS}Matrix product states}

  The previous statements can be motivated most easily in the language
  of matrix product states
  (MPS).\cite{Chen:PhysRevB.83.035107,Schuch:1010.3732v3}
  Since all its characteristics should
  be visible at zero temperature, we expect the topological phase of a
  system to be fully encoded in its ground state $|\psi\rangle$. In
  this paper we will throughout assume the absence of spontaneous symmetry
  breaking such that the ground state is unique (the more general case
  can be considered along the lines of Refs.\
  \onlinecite{Schuch:1010.3732v3,Chen:PhysRevB.84.235128}). It
  is also crucial to require an energy gap between the ground state
  and the first excited state, even in the thermodynamic limit, since
  otherwise long range correlations would exist which might spoil the
  existence of a topological invariant altogether. We regard the
  requirement of having a gap as being equivalent to demanding a
  finite correlation length.

  As is well known, any state, including the ground state
  $|\psi\rangle$, on a periodic chain of length $L$ can be represented
  as a matrix product state of the form
  \cite{Perez-Garcia:2007:MPS:2011832.2011833}
\begin{align}
  |\psi\rangle
  \ =\ \sum_{i_1,\ldots,i_L}\tr(A^{[1]\,i_1}\cdots A^{[L]\,i_L})\,|i_1\cdots i_L\rangle\ \ ,
\end{align}
  where the vectors $|i_k\rangle$ constitute an orthonormal basis of
  the Hilbert space $\cH_k$. If the dimension of the matrices
  $A^{[k]}$ remains bounded uniformly when $L$ is sent to infinity it
  makes sense to speak about the thermodynamic limit of the state
  $|\psi\rangle$. One can then specify very precise conditions under
  which the state defines correlation functions with a finite
  correlation length.
  \cite{Fannes:1990ur,Perez-Garcia:2007:MPS:2011832.2011833} At the
  same time, they ensure the existence of a mass gap even in the
  thermodynamic limit. Throughout the paper we are only interested in
  situations where $|\psi\rangle$ is finitely correlated and invariant
  under the action of $G$.

\begin{figure}
\includegraphics[]{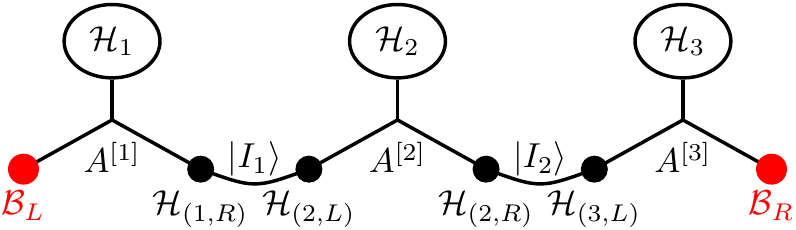}
  \caption{\label{fig:MPS}(Color online) Sketch of a matrix
    product state for a system with open boundary conditions. The
    states in the boundary spaces $\cB_L$ and $\cB_R$ (red) correspond
    to massless edge modes.}
\end{figure}

  From a mathematical perspective matrix product states arise by
  associating two auxiliary sites $(k,L)$ and $(k,R)$ to each physical
  site $k$ which carry a Hilbert space $\cH_{(k,L)}$ and
  $\cH_{(k,R)}$. Moreover we demand that
  $\cH_{(k,R)}=\cH_{(k+1,L)}^\ast$. This guarantees the existence of
  intertwiners
  $I_k:\Complex\to\cH_{(k,R)}\otimes\cH_{(k+1,L)}$. Alternatively, one
  has a state $I_k(1)=|I_k\rangle\in\cH_{(k,R)}\otimes\cH_{(k+1,L)}$,
  a completely entangled pair, which is
  invariant under the action of $G$. Under these prerequisites, the
  matrices $A^{[k]}$ can be regarded as intertwiners from
  $\cH_{(k,L)}\otimes\cH_{(k,R)}$ to $\cH_k$. The
  state $|\psi\rangle$ can then be viewed as the image of a product
  $|I\rangle=|I_1\rangle\otimes\cdots\otimes|I_{L-1}\rangle$ of
  completely entangled pairs under the map
  $A^{[1]}\otimes\cdots\otimes A^{[L]}$. By construction, the
  state $|\psi\rangle$ is invariant under the action of $G$. The
  construction of a matrix product state is sketched in Figure
  \ref{fig:MPS}.
  
  Let $R^{[k]}:G\to U(\cH_k)$ be a unitary representation of $G$ on
  $\cH_k$ and let, similarly, $D^{[k]}:G\to U(\cH_{(k,L)})$ be a
  unitary (potentially projective) representation on
  $\cH_{(k,L)}$. The intertwining property for the homomorphisms
  $A^{[k]}$ translates into the equation (see also Ref.\ 
  \onlinecite{Sanz:PhysRevA.79.042308})
\begin{align}
  \label{eq:Intertwiner}
  R^{[k]}(g)\cdot A^{[k]}
  \ =\ D^{[k]}(g)A^{[k]}D^{[k+1]}(g)^{-1}\ \ .
\end{align}
  In this equation, the maps $A^{[k]}$ are interpreted as
  homomorphisms from $\cH_{(k,R)}=\cH_{(k+1,L)}^\ast$ to $\cH_{(k,L)}$
  with values in $\cH_k$.
  It should be emphasized that the auxiliary space
  $\cH_{(k,L)}\otimes\cH_{(k,R)}$ can always be regarded as a
  representation of $G$ even when the two auxiliary spaces
  $\cH_{(k,L)}$ and $\cH_{(k,R)}$ themselves are only projective
  representations of $G$ (as long as their projective class sums up to
  the trivial one). This is due to the fact that potential phase
  factors arising in the multiplication law \eqref{eq:ProjRep} are
  canceling out on the right hand side of eq.\
  \eqref{eq:Intertwiner}.

  In a chain with open boundary conditions, the auxiliary spaces
  $\cB_L=\cH_{(1,L)}$ and $\cB_R=\cH_{(L,R)}$ at the two boundaries
  are associated with the massless edge modes and, as advertised
  before, these are capable of carrying a projective representation of
  $G$. This is equivalent to the statement that they carry a linear
  representation of two centrally extended groups
  $\tilde{G}(\Omega)$ and $\tilde{G}(-\Omega)$,
  respectively (if the system is not in a superposition of topological
  phases). The situation is pictured in Figure \ref{fig:EdgeModes}.

  It was the remarkable insight of Refs.\
  \onlinecite{Chen:PhysRevB.83.035107,Schuch:1010.3732v3}
  that the (discrete) projective class $\Omega$ is invariant under
  continuous deformations of the physical system. For this reason, it
  can be viewed as a quantitative measure for the topological phase
  the system resides in. The continuity of the deformation is
  equivalent to the preservation of a gap. Moreover, it is important
  to emphasize that the previous classification only holds as long as
  we restrict ourselves to deformations which retain the full original
  symmetry group $G$.

  If we view the same system from the angle of a different symmetry
  $G'$ and if we allow for deformations which preserve $G'$ instead of
  $G$, the classification of topological phases will change. In
  particular, one and the same system can belong to different
  topological classes, depending on the symmetry group under
  consideration. It is thus incorrect to think about $\Omega$ as being
  an inherent property of the physical system, without specifying the
  precise symmetry group the classification refers to. This basic but
  important observation will play a key role in Section
  \ref{sc:Physics}.\footnote{
  For the $S=1$ spin chain in the Haldane phase the previous
  paragraph amounts to the following statements: The system is
  invariant under $SU(2)$ as well as under $SO(3)$. If one restricts
  oneself to deformations which preserve the $SO(3)$ symmetry, the
  system is in a topologically non-trivial phase. However, it is in
  the topologically trivial phase with regard to deformations which
  only need to preserve $SU(2)$. An intuitive way of understanding
  this is the observation that the emerging $S=1/2$ boundary spins
  are not stable against the explicit addition of extra $S=1/2$
  boundary spins. On the other hand, this addition is not allowed if
  we wish to preserve $SO(3)$ invariance locally since $S=1/2$ is a
  representation of $SU(2)$ but not of $SO(3)$.
}

  It should finally be noted that the dimension of the spaces $\cB_L$
  and $\cB_R$ alone is not sufficient to discriminate between
  topological phases.
  \cite{Tu:PhysRevB.80.014401,Pollmann:2012PhRvB..86l5441P,Duivenvoorden:2012}
  It really requires knowledge of the full representation type $\Omega\in
  H^2\bigl(G,U(1)\bigr)$. In principle, the latter should be
  measurable by a suitable non-local order parameter.
  \cite{Haegeman:1201.4174v1,Pollmann:2012PhRvB..86l5441P,Duivenvoorden:2012}
  In contrast, it is not clear to us whether this knowledge can be
  inferred unambiguously from (a non-specialized version of) the
  entanglement spectrum.

  As we have just reviewed, the general principles leading to the
  classification of symmetry protected topological phases are well
  known. What is currently still missing is an explicit evaluation of
  the cohomology groups $H^2\bigl(G,U(1)\bigr)$ for general continuous
  groups $G$. Moreover, for the purpose of constructing non-trivial
  topological phases it will be important to have an explicit map
  between the boundary representations $\cB_L$ and $\cB_R$ and their
  associated projective classes $\Omega$ and $-\Omega$. Section
  \ref{sc:TopPhases} below will provide a complete solution to both
  problems. However, before we can state our results we first need to
  recall some facts about the structure of continuous groups.

\subsection[Review of $SU(2)$ spin chains: The difference between
$SU(2)$ and $SO(3)$]{\label{sc:SUReview}Review of \texorpdfstring{\textit{SU}(2)}{SU(2)}
  spin chains: The difference between \texorpdfstring{\textit{SU}(2)}{SU(2)} and
  \texorpdfstring{\textit{SO}(3)}{SO(3)}}

  In an $SU(2)$ spin chain, the spin operators $\vec{S}_k$ on each
  site take values in the spin algebra $su(2)$. The relevant
  irreducible representations are labeled by the spin
  $S\in\{0,1/2,1,3/2,\ldots\}$. By definition, the
  spin chain possesses an $SU(2)$ symmetry if the total spin generator
  $\vec{S}=\sum_k\vec{S}_k$ commutes with the Hamiltonian $H$.

  For the classification of topological phases we need to carefully
  consider which symmetry group $G$ is entering the cohomology group
  $H^2\bigl(G,U(1)\bigr)$. If the physical spins transform in
  half-integer spin representations, the group $SU(2)$ is acting
  faithfully and there is only one topological phase. Indeed, it is
  well known that $SU(2)$ only admits the trivial central extension
  $SU(2)\times U(1)$.

  The situation is different if the physical spins transform in
  integer spin representations. In that case $SU(2)$ does {\em not}
  act faithfully and the actual symmetry is only
  $SO(3)=SU(2)/\Integer_2$. However, the edge modes can transform in
  projective representations of $SO(3)$ and all of them can be thought
  of as ordinary representations of $SU(2)$. We now thus find two
  different topological classes, corresponding to edge modes
  transforming either in integer or in half-integer representations of
  $SU(2)$. \cite{Chen:PhysRevB.83.035107} The two central extensions
  (by $U(1)$) corresponding to these two classes are $SO(3)\times
  U(1)$ and $U(2)$.\footnote{Note
    that $U(2)$ is {\em not} a central extension of $SU(2)$ since
    $U(2)/U(1)=SO(3)$. Indeed, the subgroup $\Integer_2\subset SU(2)$
    is contained in $U(1)\subset U(2)$. In other words:
    $U(2)=(SU(2)\times U(1))/\Integer_2$.} It should be
  noted that the difference can already be seen in the two central
  extensions of $SO(3)$ by $\Integer_2$, namely
  $SO(3)\times\Integer_2$ and $SU(2)$.

  In view of the envisaged generalization to spin chains based on
  $SU(N)$ and other Lie groups it is useful to understand the
  difference between $SU(2)$ and $SO(3)$ more precisely in topological
  terms. When viewed as geometric manifolds, $SU(2)$ and $SO(3)$ look
  identical locally, i.e.\ they have the same underlying Lie algebra
  $su(2)$. However, they differ in their global topology. While
  $SU(2)$ is simply-connected, the group $SO(3)$ is not
  simply-connected, i.e.\ it admits non-trivial loops which cannot be
  contracted to a point. Phrased more mathematically, $SO(3)$ has
  fundamental group $\pi_1\bigl(SO(3)\bigr)=\Integer_2$ while
  $\pi_1\bigl(SU(2)\bigr)=\{1\}$. In other words, $SU(2)$ can be
  viewed as a two-fold covering of the group $SO(3)$. As we will
  review in the following subsection, the close tie between
  fundamental groups and covering groups extends to other symmetry
  groups, e.g.\ to $SU(N)$.

\subsection{\label{sc:GlobalProperties}From Lie algebras to Lie
  groups}

  Let us now consider a general spin chain whose spin operators take
  values in a Lie algebra $\g$.\footnote{The background on
    Lie theory needed for this section can for instance be found in
    \onlinecite{FultonHarris:MR1153249,Hall2003:MR1997306,Mimura:MR1122592}}
  For convenience we will assume $\g$ to be simple. The rank of $\g$
  will be denoted by $r$. The finite dimensional irreducible
  representations of $\g$ are labeled by integrable weights $\lambda$,
  i.e.\ by $r$-tuples of non-negative integers. Denote this set by
  $P^+$. Relaxing the positivity condition one obtains the weight
  lattice $P$. The root lattice will be denoted by $Q$. It is a
  sublattice of $P$ and both can be regarded as abelian groups. In
  Section \ref{sc:TopPhases} we shall show that, under certain natural
  assumptions, the topological classes of $\g$-symmetric spin chains
  are in one-to-one correspondence with the elements in the quotient
  $P/Q$.\footnote{We note the amusing fact that the order $|P/Q|$
    coincides with the determinant of the Cartan matrix of $\g$.}

\begin{figure}
\includegraphics[]{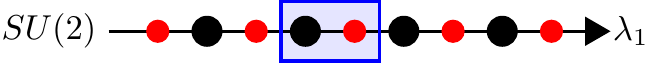}
  \caption{\label{fig:CongruenceSU}(Color online) Visualization of
    different congruence classes for $SU(2)$. The picture
    shows the weight lattice $P$ (all spins) in terms of colored
    dots. The root lattice $Q$ (integer spins) corresponds to the
    large black dots. Different colors indicate different congruence
    classes. The shaded blue box is a possible representative of
    $P/Q$.}
\end{figure}

  In the case $\g=su(2)$, the weight lattice\footnote{Recall
    that the $su(2)$ weights are twice the integral or half-integral
    magnetic quantum numbers used in physics.} is given by
  $P=\Integer$ while the root lattice is given by $Q=2\Integer$ such
  that $P/Q=\Integer_2$, see Figure \ref{fig:CongruenceSU}. This
  reproduces the classification we obtained for the symmetry group
  $SO(3)$ but not that for $SU(2)$ even though both are associated
  with the same Lie algebra $su(2)$. If at all, our assertion can
  thus only be true for a subset of symmetry groups with Lie algebra
  $\g$. In what follows we review the classification and construction
  of such Lie groups. We also single out a Lie group $PG$ which arises
  naturally from a physical perspective and whose second cohomology
  group coincides with the quotient $P/Q$.

  Any simple Lie algebra $\g$ can be exponentiated to a compact
  connected Lie group. However, as we have just seen in Section
  \ref{sc:SUReview}, several distinct Lie groups might
  have the same underlying Lie algebra $\g$. The Lie groups associated
  with $\g$ all look the same locally but they differ in their global
  topological properties, more precisely in their fundamental
  group. \cite{Mimura:MR1122592} To obtain a description of all Lie
  groups belonging to $\g$ we start with the unique simply-connected
  Lie group $G$. The Lie
  group $G$ serves as a universal cover, i.e.\ all other Lie groups
  belonging to $\g$ can be obtained by taking quotients
  $G_\Gamma=G/\Gamma$ where $\Gamma\subset\cZ(G)$ is an arbitrary
  non-trivial subgroup of the center of $G$. The groups $G_\Gamma$
  have center $\cZ(G_\Gamma)=\cZ(G)/\Gamma$ and fundamental group
  $\pi_1(G_\Gamma)=\Gamma$. It is custom to denote the centerless Lie
  group with Lie algebra $\g$ by the symbol $PG=G/\cZ(G)$ and to call
  it the projective group belonging to $G$.\footnote{Elsewhere this
    group is also called the adjoint group $\Ad(G)$
    \onlinecite{Hall2003:MR1997306}} Among the Lie groups
  associated with $\g$ it has the maximal fundamental group $\cZ(G)$,
  i.e.\ its topology is the most complicated. A list of all classical
  simple Lie algebras $\g$ and the associated simply-connected group
  $G$ can be found in Table \ref{tb:CongruenceClasses}, together with
  the relevant data for $P/Q$ and $\cZ(G)$. For readers not dealing
  with Lie theory every day we should stress that the simply-connected
  double cover of $SO(N)$ is known as $Spin(N)$.

\begin{table*}
\begin{center}
  \begin{tabular}{c|ccccccccc}
    Lie algebra $\g$ & $A_n$ & $B_n$ & $C_n$ & $D_n$ & $E_6$ & $E_7$ & $E_8$ & $F_4 $ &
    $G_2$ \\\hline\hline\\[-3mm]
    Other name & $su(n+1)$ & $so(2n+1)$ & $sp(2n)$ & $so(2n)$\\[2mm]
    $G$ & $SU(n+1)$ & $Spin(2n+1)$ & $Sp(2n)$ & $Spin(2n)$ & $E_6$ &
    $E_7$ & $E_8$ & $F_4$ & $G_2$ \\[2mm]
    $P/Q\cong\cZ(G)$ & $\Integer_{n+1}$ & $\Integer_2$ & $\Integer_2$ &
    $\Integer_4$ ($n$ odd) &
    $\Integer_3$ & $\Integer_2$ & $\{1\}$ & $\{1\}$ & $\{1\}$ \\[2mm]
    &&&&$\Integer_2\times\Integer_2$ ($n$ even)
  \end{tabular}
  \caption{\label{tb:CongruenceClasses}Simple Lie algebras $\g$ and
    their associated compact connected simply-connected Lie group
    $G$. The table also contains the congruence group $P/Q$ of $\g$
    and the center $\cZ(G)$ of $G$.}
\end{center}
\end{table*}

\subsection{\label{sc:Lifting}Lifting representations}

  In the following paragraphs we will compare the representation
  theory of the groups $G$ and $G_\Gamma$ (especially $PG$) and relate
  it to the representation theory of $\g$. By considering
  infinitesimal group actions it is clear that any finite dimensional
  representation of
  $G$, $G_\Gamma$ or $PG$ must also be a representation of $\g$. In
  contrast, the opposite conclusion only holds for the
  simply-connected Lie group $G$, the universal cover of all the
  groups $G_\Gamma$. This restriction arises from the fact that the
  center $\cZ(G)\subset G$ might act non-trivially on a
  representation, thus preventing it from descending to the quotient
  $G_\Gamma=G/\Gamma$. Nevertheless, the latter can still be regarded
  as a projective representation of $G_\Gamma$.

  In order to study this issue more systematically, let us consider an
  irreducible representation $V_\lambda$ of $\g$ (and hence $G$) with
  highest weight $\lambda\in P^+$. As a consequence of Schur's Lemma,
  the elements of the center $\cZ(G)$ are represented by multiples
  of the identity operator. Put differently, $V_\lambda$ can be viewed
  as $\dim(V_\lambda)$ copies of one and the same one-dimensional
  representation $[\lambda]$ of the abelian group
  $\cZ(G)$.\footnote{For an abelian group all irreducible
    representations are one-dimensional.} We call $[\lambda]$ the
  {\em congruence class} of $\lambda$. $[\lambda]$ can be interpreted
  as an element $[\lambda]\in\Hom\big(\cZ(G),U(1)\bigr)$ of the
  character group of $\cZ(G)$. In our situation, with $\cZ(G)$ being
  finite, the character group $\Hom\big(\cZ(G),U(1)\bigr)$ is
  isomorphic to the center $\cZ(G)$ itself, albeit the identification
  is not canonical.

  We note that the algebraic structures on $P^+$ and on
  $\Hom\big(\cZ(G),U(1)\bigr)$ (considered as an additive group) are
  compatible with the embedding specified above in the sense that
  $[\lambda+\mu]\equiv[\lambda]+[\mu]$. Indeed, the left hand side of
  this equation is determined by the action of $\cZ(G)$ on the
  irreducible representation $V_{\mu+\lambda}$. However, the latter
  can be realized as an invariant subspace of the tensor product
  $V_\lambda\otimes V_\mu$ on which the two actions of
  $\cZ(G)$ on the individual factors just multiply trivially, leading
  to the class $[\lambda]+[\mu]$. Since the trivial
  representation of $G$ is associated with the trivial representation
  $[0]$ of $\cZ(G)$, the previous relation can be used to extend the
  definition of $[\,\cdot\,]$ from $P^+$ to the full weight lattice $P$. This is
  also consistent with the observation that if $\lambda^+$ denotes
  the representation conjugate to $\lambda$, one easily finds
  $[\lambda^+]\equiv[\lambda]^+\equiv-[\lambda]$, as is implied by the
  existence of the trivial representation inside of $V_\lambda\otimes
  V_\lambda^\ast$. Moreover, all groups $G_\Gamma$ admit an action on
  $\g$ by conjugation which is insensitive to the action of the
  center. Since the generators of $\g$ can be interpreted as elements
  of $Q$, this means that the root lattice $Q$ is mapped to $[0]$ and,
  in fact, one obtains a homomorphism
  $P/Q\to\Hom\big(\cZ(G),U(1)\bigr)$. A closer inspection shows that
  the homomorphism just constructed is actually an isomorphism
  (Ref.\ \onlinecite{Hall2003:MR1997306} Theorem 8.30).\footnote{Basically
  this statement amounts to the fact that the root lattice can be
  defined as the kernel of the exponential map from $\g$ to $G$.}
  Summarizing our previous discussion, we obtain an isomorphism
\begin{align}
  \label{eq:FirstEqualityOfGroups}
  P/Q\ \cong\ \Hom\big(\cZ(G),U(1)\bigr)
     \ \cong\ \cZ(G)\ \ .
\end{align}
  Any representation $\lambda$ of $G$ with $[\lambda]\equiv[0]$ is a
  linear representation of $PG$ while all the other ones are only
  projective representations.
  
  Similar considerations apply to any subgroup $\Gamma\subset\cZ(G)$
  of the center. By the same arguments as above we can define a
  surjective homomorphism
  $[\,\cdot\,]_\Gamma:P\to\Hom(\Gamma,U(1))$. Since all the groups
  involved are abelian, one can regard the character group
  $\Hom(\Gamma,U(1))$ of $\Gamma$ as a quotient of the character group
  $\Hom\bigl(\cZ(G),U(1)\bigr)$ (see also eq.\
  \eqref{eq:EmbeddingChain}) and hence as a sublattice of $P/Q$. If
  $Q_\Gamma$ denotes the kernel of the map $[\,\cdot\,]_\Gamma$ we
  obviously obtain the isomorphisms
\begin{align}
  \label{eq:SecondEqualityOfGroups}
  P/Q_\Gamma\ \cong\ \Hom\big(\Gamma,U(1)\bigr)
     \ \cong\ \Gamma\ \ .
\end{align}
  All linear representations $\lambda$ of $G_\Gamma$ satisfy
  $[\lambda]_\Gamma\equiv[0]$. If this equation is not satisfied,
  $\lambda$ is a projective representation of $G_\Gamma$. Note that
  any representation with $[\lambda]\equiv[0]$ automatically satisfies
  $[\lambda]_\Gamma\equiv[0]$ for all $\Gamma\subset\cZ(G)$. More
  generally, the relation $[\lambda]_\Gamma\equiv[0]$ implies
  $[\lambda]_{\Gamma'}\equiv[0]$ for all
  $\Gamma'\subset\Gamma\subset\cZ(G)$. Additional details on the
  relationship between the maps $[\,\cdot\,]_\Gamma$ and
  $[\,\cdot\,]_{\Gamma'}$ for different choices of $\Gamma$ and
  $\Gamma'$ can be found in Section \ref{sc:Physics}.
  In the next section we will
  argue that all the groups appearing in eq.\
  \eqref{eq:SecondEqualityOfGroups} can also be identified with the
  cohomology group $H^2\bigl(G_\Gamma,U(1)\bigr)$, thus relating our
  findings to the classification of topological phases.

\subsection{\label{sc:CentralExtensions}Central extensions of compact
  Lie groups}

  As discussed in Section \ref{sc:MPSClass}, central extensions of
  an arbitrary group $K$ are classified by the cohomology group
  $H^2\bigl(K,U(1)\bigr)$. For a finite group $K$, the
  determination of the second cohomology group essentially reduces to
  a purely combinatorial problem. The situation is very different for
  continuous groups since now cocycles and coboundaries have to be
  measurable functions of continuous variables, resulting in an
  infinite number of constraints.

  For concreteness, we assume all Lie groups to be finite dimensional,
  compact and connected in what follows. In this case, the
  cohomology $H^2\bigl(K,U(1)\bigr)$ receives contributions
  from two sources: There might be local obstructions to the
  trivialization of cocycles. These are classified by central
  extensions of the Lie algebra belonging to $K$ and they are
  absent if $K$ is semisimple. Moreover, there might be global
  obstructions arising from the existence of non-contractible loops in
  $K$, i.e.\ from a non-trivial fundamental group
  $\pi_1(K)$.\cite{Bargmann:1954gh}

  Our previous statements can brought into a mathematically precise
  form and they result in the following proposition (for a proof see
  e.g.\ Ref.\ \onlinecite{Moore1964:MR0171880} Prop. 2.1)
\begin{proposition}
  \label{prop:Cohomology}
  Let $K$ be a finite dimensional compact connected simple Lie group;
  then there is a canonical isomorphism
\begin{align}
  \label{eq:CohomologyGeneral}
  H^2\bigl(K,U(1)\bigr)
  \ \cong\ \Hom\bigl(\pi_1(K),U(1)\bigr)\ \ .
\end{align}
\end{proposition}
  \noindent
  Since $\pi_1(K)$ is finite and abelian in the cases of interest, the
  right hand side actually consists of {\em all} representations of
  $\pi_1(K)$ and can be identified with the group $\pi_1(K)$ itself
  (even though not in a canonical way).

  Let us now discuss the implications of the previous proposition
  for simply-connected simple Lie groups $G$. Since
  the fundamental group is trivial, one immediately finds that
  $H^2\bigl(G,U(1)\bigr)$ is trivial as well. In other words, $G$ does neither
  admit non-trivial central extensions nor non-trivial projective
  representations. All finite dimensional representations of the
  underlying Lie algebra $\g$ lift to linear representations of $G$.

  In the next step we drop the simply-connectedness, i.e.\ we allow
  for non-contractible loops. As was recalled in Section
  \ref{sc:GlobalProperties}, every simple Lie group can
  be written as $G_\Gamma=G/\Gamma$ where $G$ is its simply-connected
  universal cover and $\Gamma\subset\cZ(G)$ is a subgroup of the
  center of the latter. The fundamental group of $G_\Gamma$ can be
  written as $\pi_1(G_\Gamma)=\Gamma$. In order
  to illustrate the content of Proposition \ref{prop:Cohomology}, we
  are now constructing the central extensions of $G_\Gamma$
  explicitly. Fix an element $\Omega\in H^2\bigl(G_\Gamma,U(1)\bigr)$
  and interpret it as a representation $\Omega:\Gamma\to U(1)$. The
  associated central extension is given by
\begin{align}
  \tilde{G}_\Gamma(\Omega)
  \ =\ \bigl(G\times U(1)\bigr)/\Gamma\ \ ,
\end{align}
  where the central subgroup $\Gamma\subset\cZ(G)$ of $G$ is embedded
  diagonally into $G\times U(1)$ according to the prescription
  $\gamma\mapsto\bigl(\gamma,\Omega(\gamma)\bigr)$. Our previous arguments also
  imply that the projective representations of $G_\Gamma$ are just
  the representations of $G$ (or $\g$) themselves. Different
  projective classes correspond to different actions of the subgroup
  $\Gamma$. Indeed, due to Schur's Lemma the center $\Gamma$ can
  always be interpreted as being embedded in $U(1)$ (possibly not
  injectively) when acting on an irreducible representation.

\section{\label{sc:TopPhases}Topological phases of gapped spin chains}

  In this section we will give a classification of topological phases
  in gapped spin chains whose spin operators belong to an arbitrary
  simple Lie algebra $\g$. This is achieved by evaluating the
  cohomology groups $H^2\bigl(G_\Gamma,U(1)\bigr)$ explicitly by
  relating them to the central subgroup $\Gamma\subset\cZ(G)$ defining
  $G_\Gamma$. We also provide a dictionary that characterizes massless
  boundary modes according to the congruence class of their
  representation. We conclude with a detailed application of our
  general result to each individual simple Lie group. Among these, the
  symmetry group $PSU(N)$ is the most interesting since the number of
  distinct topological phases turns out to increase with $N$. Also the
  symmetry groups $PSO(2n)$ stand out since their four topological
  phases are characterized by either $\Integer_2\times\Integer_2$ or
  $\Integer_4$, depending on whether $n$ is even or odd.

\subsection{\label{sc:GeneralCase}Topological classes for spin chains
  with general Lie group symmetry}

  In all what follows we use the notation introduced in Sections
  \ref{sc:GlobalProperties} and \ref{sc:Lifting}. We shall assume that
  the physical on-site Hilbert spaces $\cH_k$ can be regarded as
  linear representations of the group $G_\Gamma$. In particular,
  the central subgroup $\Gamma\subset\cZ(G)$ acts trivially on each
  $\cH_k$ such that these spaces are associated with the class $[0]\in
  P/Q_\Gamma$.

  We are now prepared to present the main result of the paper.
  Combining the statements of Section \ref{sc:Lifting} and of Section
  \ref{sc:CentralExtensions}, the classification of topological phases
  can be obtained from the following chain of isomorphisms,
\begin{align}\nonumber
  \label{eq:EqualityOfGroups}
  H^2\bigl(G_\Gamma,U(1)\bigr)
  \ &\cong\ \Hom\big(\Gamma,U(1)\bigr)\\
  \ &\cong\ \Gamma
  \ \cong\ P/Q_\Gamma\ \ .
\end{align}
  In other words, the different topological phases of a spin chain
  with symmetry group $G_\Gamma$ are in one-to-one correspondence with
  the elements of its fundamental group $\Gamma$. In particular, the
  topological phases of a system with $PG$ symmetry can be identified
  with the center of $G$. In this case, the previous equation reduces
  to
\begin{align}\nonumber
  \label{eq:ThirdEqualityOfGroups}
  H^2\bigl(PG,U(1)\bigr)
  \ &\cong\ \Hom\big(\cZ(G),U(1)\bigr)\\
  \ &\cong\ \cZ(G)
  \ \cong\ P/Q\ \ .
\end{align}
  The interpretation of the center as the quotient of the weight
  lattice $P$ of $\g$ modulo its root lattice $Q$ is sometimes useful
  for the concrete evaluation of $\cZ(G)$, e.g.\ for exceptional
  groups like $E_6$. More importantly, it provides the avenue for a
  characterization of topological phases in terms of edge modes as
  will be explained in Section \ref{sc:EdgeModes}. The relevant data
  for $P/Q$ (and hence $\cZ(G)$) for different choices of $\g$ can be
  found in Table \ref{tb:CongruenceClasses}. The important question
  how to determine the relevant symmetry group $G_\Gamma$ entering
  eq.\ \eqref{eq:EqualityOfGroups} will be addressed in Section
  \ref{sc:Physics}. Let us just emphasize here that one can be certain
  not to miss a possible phase if one employs eq.\
  \eqref{eq:ThirdEqualityOfGroups} instead. In this sense, the
  symmetry group $PG$ can be regarded as a kind of master symmetry.

\subsection{\label{sc:EdgeModes}Edge mode representations as an
  indicator for the topological phase}

  We will argue in Section \ref{sc:Physics} that the topological
  phases of systems with $G_\Gamma$-symmetry admit, in many cases, a
  natural embedding into the topological phases of systems with
  $PG$-symmetry. Hence we will restrict the following analysis to the
  symmetry group $PG$.
  
  Let us thus consider a $PG$-symmetric gapped spin chain with a
  unique $PG$-invariant ground state which resides in a well-defined
  topological class. According to our previous discussions this
  statement has three implications. Firstly, all irreducible
  representations $\lambda$ appearing in the decomposition
\begin{align}
  \label{eq:Decomposition}
  \cH_k\ =\ \bigoplus_\lambda V_\lambda
\end{align}
  of the physical on-site Hilbert spaces $\cH_k$ should belong to the
  trivial class $[0]\in P/Q$. Secondly, there should exist a unique
  class $\Omega\in\Hom\bigl(\cZ(G),U(1)\bigr)$ labeling the
  topological phase.\footnote{This condition can be violated if the boundary
    Hilbert space $\cB$ describing the gapless edge mode at one
    boundary is not an irreducible representation.} Thirdly, the edge modes
  (possibly virtual) on the left hand side and on the right hand side
  of the (reduced) system should transform in representations which
  correspond to the projective classes $\Omega\in P/Q$ and $-\Omega\in
  P/Q$, respectively.\footnote{The assignment of $\Omega$ to the left
    edge is by convention only and represents no loss of
    generality. We could equally well flip the orientation of the
    whole system. In any case, the edges need to transform in opposite
    classes since the action  of the symmetry group $PG$ on the total
    chain should correspond to the trivial class $[0]$, i.e.\ to a
    linear representation.} If we decompose the auxiliary Hilbert
  space $\cB_L=\cH_{(1,L)}$ (or $\cB_R=\cH_{(L,R)}$) at the boundary into
  irreducible representations of $\g$ similar to eq.\
  \eqref{eq:Decomposition}, then all the $\lambda$ should belong to
  the same class $\Omega\in P/Q$ (or $-\Omega\in P/Q$). The previous
  few lines clearly exhibit the need for an efficient way of
  determining the projective class of a given representation
  $\lambda$ of $\g$.

\begin{table}
\begin{center}
  \begin{tabular}{ccc}
    Lie algebra & Congruence vector(s) $\nu$ & Modulus $M$\\\hline\hline\\[-3mm]
    $A_n$ & $(1,2,\ldots,n)$ & $n+1$ \\[2mm]
    $B_n$ & $(0,\ldots,0,1)$ & $2$ \\[2mm]
    $C_n$ & $(1,0,1,0,\ldots)$ & $2$ \\[2mm]
    $D_{2n+1}$ & $(0,\ldots,0,1,1)$ & $2$ \\[2mm]
      & $(2,0,2,\ldots,2,2n-1,2n+1)$ & $4$ \\[2mm]
    $D_{2n}$ & $(0,\ldots,0,1,1)$ & $2$ \\[2mm]
      & $(2,0,2,\ldots,2,0,2n-2,2n)$ & $4$ \\[2mm]
    $E_6$ & $(1,-1,0,1,-1,0)$ & $3$ \\[2mm]
    $E_7$ & $(0,0,0,1,0,1,1)$ & $2$
  \end{tabular}
\end{center}
  \caption{\label{tab:CongVec}Congruence vectors for simple Lie
    algebras.\cite{Lemire:1979qd}}
\end{table}

  Fortunately, there exists an explicit formula which determines the
  congruence classes $[\lambda]\in P/Q$ of any irreducible
  representation $\lambda$ of $\g$. \cite{Lemire:1979qd} If
  $\lambda=(\lambda_1,\ldots,\lambda_r)\in P^+$ denotes the associated
  integrable weight, one simply finds 
\begin{align}
  \label{eq:ProjClass}
  [\lambda]\ \equiv\ \sum_{i=1}^r\lambda_i\nu_i\mod M\ \ ,
\end{align}
  where the congruence vectors $\nu$ are summarized in Table
  \ref{tab:CongVec}. In all cases but $so(4n)$ ($=D_{2n}$) the class
  $[\lambda]$ is specified by a single number. Only for $so(4n)$ there
  are two choices of $(\nu,M)$ one has to consider at the same
  time. In this case, the class is given by a tuple $[a,b]$ of two
  numbers. Since formula \eqref{eq:ProjClass} is pretty abstract, we
  will use the subsequent sections to evaluate it in great detail for
  all relevant groups. We shall begin with $SU(N)$ and continue with
  all the remaining simple simply-connected Lie groups, including
  $Spin(N)$ (the two-fold cover of $SO(N)$) and $SP(2N)$ as well as
  the exceptional groups $E_6$ and $E_7$. The remaining exceptional
  Lie groups $E_8$, $F_4$ and $G_2$ have a trivial center and hence do
  not allow for non-trivial topological phases.

\subsection[Topological classes for $SU(N)$ spin
chains]{\label{sc:SUCase}Topological classes for \texorpdfstring{\textit{SU(N)}}{SU(N)} spin
  chains}

  We assume that $N\geq2$ since $SU(1)$ is just the trivial group.
  The group $SU(N)$ is simply-connected and it has a center
  $\Integer_N$. When defined in matrix form, the center consists of
  the matrices $\omega^l\idop$ with $\omega=\exp(2\pi i/N)$ and
  $l=0,\ldots,N-1$. The restriction of the prefactor to the $N$
  distinct $N^{\text{th}}$ roots of unity is implied by the
  requirement that $SU(N)$ matrices should have unit determinant.

  The group $SU(N)$ serves as the universal cover of the projective
  special unitary group $PSU(N)=SU(N)/\Integer_N$. According to our
  general result \eqref{eq:ThirdEqualityOfGroups}, topological phases
  of $SU(N)$ spin chains are classified by the cohomology group
\begin{align}
  H^2\bigl(PSU(N),U(1)\bigr)
  \ \cong\ \Integer_N\ \ .
\end{align}
  In other words there are $N$ distinct topological phases. For $N=2$
  this reproduces the familiar result for $PSU(2)=SO(3)$ (see also
  Sections \ref{sc:SUReview} and \ref{sc:SOCase}).

  Let us now describe which type of edge mode indicates the presence
  of which topological phase. As explained in Section
  \ref{sc:EdgeModes} this requires knowledge about the congruence
  class of all irreducible representations of $SU(N)$. Representations
  of $SU(N)$ can be described in terms of integrable weights
  $\lambda=(\lambda_1,\ldots,\lambda_{N-1})$ as above or,
  alternatively, in terms of Young tableaux
  $\lambda=\{l_1;\ldots;l_{N-1}\}$. In terms of the weight, the
  partition of the associated Young tableau is specified by the
  numbers
\begin{align}
  \label{eq:Partition}
  l_i\ =\ \sum_{k=i}^{N-1}\lambda_k\ \ .
\end{align}
  By definition, the number $l_i$ determines the number of boxes in
  the $i^{\text{th}}$ row of the tableau.

\begin{figure}
\includegraphics[scale=0.7]{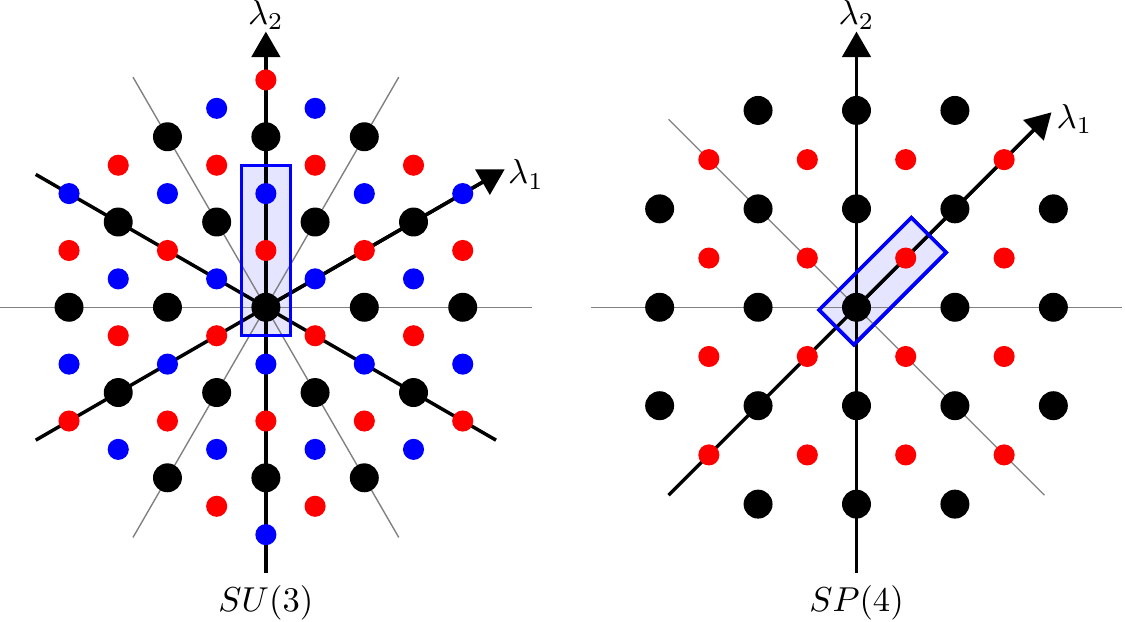}
  \caption{\label{fig:CongruenceSUSP}(Color online) Visualization of
    different congruence classes for $SU(3)$ and $SP(4)$. The pictures
    show the respective weight lattice $P$ in terms of colored
    dots. The root lattice $Q$ corresponds to the large black
    dots. Different colors indicate different congruence
    classes. The shaded blue boxes are possible representatives of
    $P/Q$. We clearly see that, for $SP(4)$, the topological class is
    independent of $\lambda_2$.}
\end{figure}

  According to our general result \eqref{eq:ProjClass} and Table
  \ref{tab:CongVec}, the projective class of a representation
  $\lambda$ is given by
\begin{align}
  \label{eq:SUClass}
  [\lambda]
  \ \equiv\ \sum_{k=1}^{N-1}k\lambda_k\mod N\ \ .
\end{align}
  This formula divides the weight lattice $P$ into $N$ sublattices,
  each of them labeled by an element of $P/Q$. An illustration of this
  fact is shown in Figure \ref{fig:CongruenceSU} and in Figure
  \ref{fig:CongruenceSUSP} for the particular cases of $SU(2)$ and
  $SU(3)$, respectively.

  We will now briefly recall in which way the $N$ different classes of
  $SU(N)$ representations correspond to the $N$ different
  representations of the center $\Integer_N\subset SU(N)$. If
  $\rho:SU(N)\to U(V_\lambda)$ denotes the irreducible
  representation with highest weight $\lambda$, the center will act as
  follows,
\begin{align}
  \rho(\omega^l\idop)
  \ =\ \omega^{l[\lambda]}\idop\ \ .
\end{align}
  This equation is evident for the trivial representation and for the
  fundamental representation $\lambda=(1,0,\ldots,0)$ (which has
  $[\lambda]\equiv1$ and can thus be regarded as the generator of
  $\Integer_N$). The general validity follows from linear
  extrapolation (i.e.\ from taking multiple tensor products of the
  fundamental representation).

  We wish to emphasize that formula \eqref{eq:SUClass} admits a nice
  interpretation in terms of Young tableaux: The projective class of a
  representation $\lambda$ just corresponds to the number of boxes
  $|\lambda|$ modulo $N$. Indeed, a simple rewriting of eq.\
  \eqref{eq:SUClass} using the identity \eqref{eq:Partition} yields
\begin{align}
  \label{eq:SUClassYoung}
  [\lambda]
  \ \equiv\ \sum_{i=1}^{N-1}l_i\mod N
  \ \equiv\ |\lambda|\mod N\ \ ,
\end{align}
  This result can also be understood as follows. The basic
  representation of $SU(N)$ is the $N$-dimensional fundamental
  representation. It is represented by a Young tableau with a single
  box. Hence it has $[\lambda]\equiv1$ and can be regarded as the
  generator of the group $\Integer_N$. All the other representations
  of $SU(N)$ can be found in iterated tensor product of the
  fundamental representation with itself. By the Littlewood-Richardson
  rule for calculating tensor products, the number of boxes (and hence
  the projective class) increases by one unit in each iteration until
  we eventually reach the $N^{\text{th}}$ power of the tensor
  product. Here, the phase is reset to zero
  and the counting starts anew. In the process of calculating tensor
  products one might need to delete columns with $N$ boxes. However,
  deleting $N$ boxes does not have an effect if the number of boxes is
  only counted modulo $N$ anyway.

\subsection[Topological classes for $Spin(N)$ spin
chains]{\label{sc:SOCase}Topological classes for \texorpdfstring{\textit{Spin(N)}}{Spin(N)}
  spin chains}

  Let us now look at the orthogonal symmetry groups $SO(N)$.
  In what follows we restrict our attention to $N\geq3$ since
  $SO(1)=\Integer_2$ is discrete and $SO(2)=U(1)$ fails to be
  simple. Since $SO(N)$ is not simply-connected, it is more
  appropriate for the purpose of our paper to speak about the
  universal covering group $Spin(N)$ which is a two-fold cover of
  $SO(N)$. As usual, the covering implies the identity
  $SO(N)=Spin(N)/\Integer_2$. For $N=3$ we recover the familiar case
  $Spin(3)=SU(2)$ with $SO(3)=Spin(3)/\Integer_2$.

  Surprisingly, the groups $Spin(N)$ fall into two (actually three)
  separate families with rather different properties as can be
  inferred from Table \ref{tb:CongruenceClasses}. For odd $N=2n+1$
  ($n\geq1$), the center is $\Integer_2$ while for even $N=2n$ the
  center is $\Integer_4$ for odd values of $n$ and
  $\Integer_2\times\Integer_2$ for even $n$. The cohomology groups
  classifying the topological phases of $Spin(N)$ symmetric spin
  chains are thus given by
\begin{align}\nonumber
  H^2\bigl(SO(2n+1),U(1)\bigr)
  \ &\cong\ \Integer_2
  \quad\quad\quad\quad\text{ and }\quad\\
  H^2\bigl(SO(2n)/\Integer_2,U(1)\bigr)
  \ &\cong\ \begin{cases}
         \Integer_4&,\ n\text{ odd}\\[2mm]
         \Integer_2\oplus\Integer_2&,\ n\text{ even}\ \ .
       \end{cases}
\end{align}
  In particular, there are four phases if $N$ is even and two phases
  if $N$ is odd. We will treat these two cases separately in what
  follows. A partial classification, focusing on $SO(N)$, has
  previously appeared in Ref.\ \onlinecite{Haegeman:1201.4174v1}.

\subsubsection[Case $Spin(2n+1)$]{Case \texorpdfstring{\textit{Spin}$(n+1)$}{Spin(n+1)}}

  For odd $N=2n+1$ ($n\geq1$), the center is $\Integer_2$ and there
  are two different classes of representations. They can be
  distinguished by the last entry of the Dynkin label
  $\lambda=(\lambda_1,\ldots,\lambda_n)$,
\begin{align}
  [\lambda]
  \ \equiv\ \lambda_n\mod 2\ \ .
\end{align}
  If $\gamma$ is the generator of $\Integer_2\subset Spin(N)$ and
  $\rho:Spin(N)\to U(V_\lambda)$ denotes the irreducible
  representation with highest weight $\lambda$, the center is
  represented by
\begin{align}
  \label{eq:CenterSpinOdd}
  \rho(\gamma)\ =\ (-1)^{[\lambda]}\idop\ \ .
\end{align}
  Accordingly, the situation is very similar to that of
  $SU(2)$. Representations with $[\lambda]\equiv0$ are linear
  representations of $Spin(N)$ and of $SO(N)$. On the other hand,
  representations with $[\lambda]\equiv1$ are spinorial, i.e.\ they
  are linear representations of $Spin(N)$ but only projective ones of
  $SO(N)$. Since the center of $SO(N)$ is trivial for $N=2n+1$, this
  covers all possible cases.

\subsubsection[Case $Spin(2n)$]{Case \texorpdfstring{\textit{Spin}$(2n)$}{Spin(2n)}}

  The treatment of $SO(N)$ with even $N=2n$ ($n\geq2$) becomes
  slightly more involved but also more interesting. In this case, the
  center of $Spin(N)$ is $\Integer_2\times\Integer_2$ for even $n$ and
  $\Integer_4$ for odd $n$.\footnote{We shall write the center
    additively below for better comparison with $P/Q$.} This observation
  in particular implies that the groups $SO(N)=Spin(N)/\Integer_2$
  have a center $\Integer_2$ themselves such that one also needs to
  consider the group $PSO(2n)=SO(2n)/\Integer_2$.\footnote{For odd $n$
  there exist three choices of subgroups
  $\Integer_2\subset\Integer_2\times\Integer_2$. One of them leads to
  $SO(2n)$ while the other two lead to isomorphic groups which are
  known as {\em semispinor} groups. \cite{Mimura:MR1122592} The
  semispinor groups $SS(2n)$ are isomorphic 
  to $SO(2n)$ for $n=4$ but not for $n>4$.} In order to determine the
  class of a representation $\lambda=(\lambda_1,\ldots,\lambda_n)$ we
  have to calculate the $\Integer_2\oplus\Integer_4$-valued quantity
\begin{align}
  \label{eq:PhasesSpinEven}
  &[\lambda]
  \ =\ \left[\begin{matrix}[\lambda]_1\\ [\lambda]_2\end{matrix}\right]
  \ \equiv \\ &  \left[\begin{array}{cr}\lambda_{n-1}+\lambda_n&\mod2\\
            2\lambda_1+2\lambda_3+\cdots+(n-2)\lambda_{n-1}+n\lambda_n&\mod4\end{array}\right].\nonumber
\end{align}
  The first entry $[\lambda]_1$ determines whether the representation
  is a linear representation of $SO(2n)$ ($[\lambda]_1\equiv0$) or
  rather a projective one ($[\lambda]_1\equiv1$). The second entry
  $[\lambda]_2$ is required to produce the correct group structure of
  $\cZ\bigl(Spin(2n)\bigr)$ and it is relevant when it comes to
  determining whether $\lambda$ is a representation of $PSO(2n)$. For
  simplicity of presentation, we shall treat the cases $n$ even and
  $n$ odd separately.
  
  We start with $n$ even. Note that the second entry $[\lambda]_2$ is
  always even in this case. Moreover, both components of $[\lambda]$
  are completely independent. Hence precisely four of the eight
  possibilities,
\begin{align}
  [0,0],\ [0,2],\ [1,0],\ [1,2]\ \ ,
\end{align}
  are realized and one can easily check that they satisfy an addition
  law corresponding to $\Integer_2\oplus\Integer_2$ (considered as a
  subgroup of $\Integer_2\oplus\Integer_4$). If $\gamma=[1,0]$ and
  $\epsilon=[0,2]$ denote the generators of these two central
  subgroups $\Integer_2\subset Spin(2n)$, their action on an
  irreducible representation $\rho:Spin(2n)\to U(V_\lambda)$ of
  highest weight $\lambda$ is given by
\begin{align}
  \rho(\gamma)
  \ =\ (-1)^{[\lambda]_1}\idop
  \quad\text{ and }\quad
  \rho(\epsilon)
  \ =\ e^{\frac{i\pi}{2}[\lambda]_2}\idop\ \ .
\end{align}
  Representations $\lambda$ of $Spin(2n)$ with $[\lambda]=[0,0]$ are
  linear representations of $PSO(2n)$. All the remaining ones
  correspond to projective representations of $PSO(2n)$.
    
  If we turn to $n$ odd, the analysis becomes even simpler. Now the
  two entries $[\lambda]_1$ and $[\lambda]_2$ of $[\lambda]$ are
  either both even or both odd. Put differently, the first
  component $[\lambda]_1$ is completely determined by the second
  $[\lambda]_2$ by taking its value modulo two. This again realizes
  four of the eight possibilities,
\begin{align}
  [0,0],\ [1,1],\ [0,2],\ [1,3]\ \ ,
\end{align}
  but now with an addition law corresponding to $\Integer_4$ (again
  considered as a subgroup of $\Integer_2\oplus\Integer_4$), the
  generator being $\eta=[1,1]$. On an irreducible
  representation $\rho:Spin(2n)\to U(V_\lambda)$ of highest weight
  $\lambda$, the center acts as
\begin{align}
  \rho(\eta)
  \ =\ e^{\frac{i\pi}{2}[\lambda]_2}\idop\ \ .
\end{align}
  The generator $\eta^2$ of the subgroup
  $\Integer_2\subset\Integer_4\subset Spin(2n)$ which needs to be used
  to descend from $Spin(2n)$ to $SO(2n)$ is mapped to $\pm\idop$ under
  $\rho$, depending on whether $[\lambda]_2$ is even or odd. We thus
  obtain the following three level hierarchy: Representations of
  $Spin(2n)$ with $[\lambda]_2\equiv0$ are linear representations of
  $SO(2n)$ and $PSO(2n)$. If $[\lambda]_2\equiv2$ one deals with a
  linear representation of $SO(2n)$ which is only a projective
  representation of $PSO(2n)$. And in the two remaining cases, one has
  a projective representation of $SO(2n)$ and $PSO(2n)$.

  We note that in both of the superordinate cases treated, even and
  odd $n$, there exist modifications of formula
  \eqref{eq:PhasesSpinEven} which give the classification of
  topological phases in a more direct and canonical way -- in the
  first case one could divide the second component by two and in the
  second case one could restrict the attention to the second component
  from the very beginning. We decided to present both cases on the
  same footing in order to stay close to the original reference.
  \cite{Lemire:1979qd} It seems plausible that our results also have
  a natural explanation in terms of Young tableaux. However, in this
  paper we refrain from adopting this perspective.

\subsection[Topological classes for $SP(2N)$ spin
chains]{\label{sc:SPCase}Topological classes for \texorpdfstring{\textit{SP(}2\textit{N)}}{SP(2N)}
  spin chains}

  The group $SP(2N)$ is simply-connected and its center is isomorphic
  to $\Integer_2$. We should carefully note that there we
  are talking about the {\em compact} symplectic group $SP(2N)$ of
  rank $N$ (see below for a brief comment on the non-compact
  version). As usual, the topological phases are classified by the
  cohomology group
\begin{align}
  H^2\bigl(SP(2N)/\Integer_2,U(1)\bigr)
  \ \cong\ \Integer_2\ \ .
\end{align}
  We thus have two distinct topological phases. Given any weight
  $\lambda=(\lambda_1,\ldots,\lambda_N)$, the associated congruence
  class is determined by the number\footnote{For $N=2$ the second
    contribution involving $\lambda_3$ is omitted.}
\begin{align}
  [\lambda]\ \equiv\ \lambda_1+\lambda_3\mod2\ \ .
\end{align}
  The two different values of $[\lambda]$ divide the weight lattice
  $P$ into two sublattices. For $SP(4)$ this is depicted in Figure
  \ref{fig:CongruenceSUSP}. In an irreducible representation
  $V_\lambda$ of highest weight $\lambda$, the center
  $\Integer_2\subset SP(2N)$ is implemented in the same fashion as in
  eq.\ \eqref{eq:CenterSpinOdd}. Representations with
  $[\lambda]\equiv0$ are representations of $SP(2N)$ and
  $SP(2N)/\Integer_2$ while $[\lambda]\equiv1$ leads to linear
  representations of $SP(2N)$ which are projective representations of
  $SP(2N)/\Integer_2$.

  In order to prevent potential confusion, let us finally comment on
  the (probably more familiar) {\em non-compact} group
  $SP(2N,\Real)$. This group arises as the symmetry group of a
  symplectic form defined on a $2N$-dimensional real vector space. The
  fundamental group of $SP(2N,\Real)$ is given by
  $\pi_1\bigl(SP(2N,\Real)\bigr)=\Integer$. In order to arrive at a
  simply-connected group one thus needs to pass on to an infinite
  cover of $SP(2N,\Real)$. The group also has a well-known double
  cover, the so-called metaplectic group. From a representation
  theoretic point of view, the transition from the compact instance of
  a group to a non-compact version requires one to replace finite
  dimensional representations with infinite dimensional ones, just
  alone for reasons of unitarity. The topological classification of
  systems involving infinite dimensional representations is beyond the
  scope of this paper. However, our example shows that one needs to be
  very precise about the real form and the global structure of the
  symmetry group under consideration.

\subsection[Topological classes for $E_6$ and $E_7$ spin
chains]{\label{sc:ECase}Topological classes for \texorpdfstring{\textit{E}$_6$}{E(6)} and
  \texorpdfstring{\textit{E}$_7$}{E(7)} spin chains}

  Just for completeness we also treat the two exceptional cases in the
  $E$-series. By abuse of notation we also use the symbols $E_6$ and
  $E_7$ for the simply-connected groups associated with the
  corresponding Lie algebras. From Table \ref{tb:CongruenceClasses} we
  infer that the respective centers of these groups are given by
  $\Integer_3$ and $\Integer_2$. We immediately conclude that the
  cohomology groups classifying the topological phases are given by
\begin{align}\nonumber
  H^2\bigl(E_6/\Integer_3,U(1)\bigr)
  \ &\cong\ \Integer_3
  \qquad\text{ and }\\\qquad
  H^2\bigl(E_7/\Integer_2,U(1)\bigr)
  \ &\cong\ \Integer_2\ \ .
\end{align}
  Hence there are three topological phases of $E_6$-invariant and two
  phases of $E_7$-invariant spin chains.

  Let us discuss the $E_6$ case first. The representations
  $(\lambda_1,\ldots,\lambda_6)$ of $E_6$ fall into three different
  classes according to the value of
\begin{align}
  [\lambda]
  \ \equiv\ \lambda_1-\lambda_2+\lambda_4-\lambda_5\mod3\ \ .
\end{align}
  If $\gamma\in\Integer_3\subset E_6$ is the generator of the center,
  the action in an irreducible representation $\rho:E_6\to
  U(V_\lambda)$ of highest weight $\lambda$ is determined by
\begin{align}
  \rho(\gamma)\ =\ e^{\frac{2\pi i}{3}[\lambda]}\idop\ \ .
\end{align}
  Representations with $[\lambda]\equiv0$ are linear representations
  of the projective group $E_6/\Integer_3$. The remaining two classes
  are only linear representations of $E_6$ but projective
  representations of $E_6/\Integer_3$.

  Let us now turn our attention to $E_7$. The representations
  $(\lambda_1,\ldots,\lambda_7)$ of $E_7$ fall into two different
  classes according to the value of
\begin{align}
  [\lambda]
  \ \equiv\ \lambda_4+\lambda_6+\lambda_7\mod2\ \ .
\end{align}
  The action of the generator $\gamma\in\Integer_2\subset E_7$ on an
  irreducible representation of highest weight $\lambda$ is specified
  by formula \eqref{eq:CenterSpinOdd}. Representations with
  $[\lambda]\equiv0$ are linear representations of $E_7$ and
  $E_7/\Integer_2$. In contrast, representations with
  $[\lambda]\equiv1$ are linear representations of $E_7$ but only
  projective representations of $E_7/\Integer_2$.

\section{\label{sc:Physics}Physical perspectives}

  In Section \ref{sc:TopPhases} we classified topological phases for
  all spin chains whose spins belong to a simple Lie algebra $\g$. The
  classification was intimately related to a division of
  representations of $\g$ -- thought of as becoming manifest in
  gapless edge modes -- into different classes of projective
  representations of a Lie group $G_\Gamma$ associated with $\g$. In
  this section we will analyze which of the possible Lie groups
  $G_\Gamma$ is actually the relevant symmetry. We will also
  investigate the hierarchy of topological phases that arises by
  considering one and the same system from different perspectives,
  based on symmetries $G_\Gamma$ and $G_{\Gamma'}$ where $\Gamma$ and
  $\Gamma'$ are related by the inclusion
  $\Gamma'\subset\Gamma\subset\cZ(G)$. Moreover, we point out
  an interesting connection of our results with a natural
  generalization of Haldane's conjecture to arbitrary spin chains. In
  the final part of this section we illustrate our general
  considerations with two examples.

\subsection{\label{sc:SymmetryGroup}Identification of the symmetry
  group}

  In the following we will consider a fixed gapped spin system with
  spin operators in a simple Lie algebra $\g$ and a Hamiltonian that
  commutes with all elements of $\g$. Furthermore, we assume the
  action of $\g$ on the total Hilbert space to be faithful and the
  existence of a unique and $\g$-invariant ground state. The precise
  symmetry group which is relevant for the classification of potential
  topological phases, see eq.\ \eqref{eq:EqualityOfGroups}, depends on
  the nature of the on-site Hilbert spaces $\cH_k$.\footnote{We note
    in passing the following
    important facts: When the symmetries are restricted to groups of
    type $G_\Gamma$, i.e.\ as long as potential space-time symmetries
    and internal symmetries are disregarded, what matters are really
    only the Hilbert spaces under consideration. As an operator, the
    Hamiltonian $H$ itself will transform as $H\mapsto gHg^{-1}$ and
    will be insensitive to the action of the center $\cZ(G)\subset
    G$. Put differently, if $H$ commutes with all elements of the Lie
    algebra $\g$, then it will be invariant under the action of all
    elements $g\in G_\Gamma$ for any choice of
    $\Gamma\subset\cZ(G)$. This simple observation has its origin in
    Schur's Lemma and in the fact that irreducible representations of
    $G_\Gamma$ are a subset of irreducible representations of
    $\g$. Similarly, if the ground state is $\g$-invariant it will
    also automatically be $G_\Gamma$-invariant.} The simply-connected
  Lie group $G$ can always be regarded as a symmetry of the
  system. However, its action on the Hilbert spaces $\cH_k$ might not
  be faithful, leading to the existence of non-trivial kernels
  $\Gamma_k$. Whenever $\g$ acts faithfully on the total Hilbert space
  this kernel will be a subgroup $\Gamma_k\subset\cZ(G)$ of the center
  of $G$. Under these circumstances, the {\em actual} symmetry group
  (neglecting symmetries not related to $\g$) is $G_A=G/\Gamma_A$,
  with $\Gamma_A=\cap_k\Gamma_k$ being the intersection of all kernels
  $\Gamma_k$, and it is {\em this} group which enters the calculation
  of the cohomology group \eqref{eq:EqualityOfGroups} that
  characterizes potential topological phases. Note that the actual
  symmetry group as defined above might (and will generally) differ
  from that obtained by identifying $\Gamma$ with the kernel of $G$
  that arises when acting on the total Hilbert space
  $\cH=\bigotimes_k\cH_k$.\footnote{An $SU(2)$-invariant chain with
    $S=1/2$ and an even number of sites provides a simple example for
    this statement.} It is thus important to distinguish between the
  overall symmetry and symmetries that are realized locally -- even in
  the absence of translation invariance.

  Our previous statements can easily be connected to our discussion of
  congruence classes of representations of $G$ in Section
  \ref{sc:TopPhases}. The system has symmetry $G_\Gamma=G/\Gamma$ if
  {\em all} physical on-site Hilbert spaces $\cH_k$ are linear
  representations of $G_\Gamma$, i.e.\ if
  $[\cH_k]_\Gamma\equiv[0]$. In contrast, it is {\em not} required
  that {\em all} these representations are faithful. Instead we are
  searching for the ``smallest'' among the groups $G_\Gamma$ which is
  still linearly represented on {\em all} spaces $\cH_k$. Inverting
  the logic, the {\em actual} symmetry group $G_A=G/\Gamma_A$ of the
  system is associated with the maximal subgroup
  $\Gamma_A\subset\cZ(G)$ such that $[\cH_k]_{\Gamma_A}\equiv[0]$.

\subsection{\label{sc:Hierarchies}Hierarchies of topological phases}

  As a physical system can be invariant under more than one of the
  groups $G_\Gamma$ it seems appropriate to discuss the relation
  between the potential topological phases predicted for different
  choices of $\Gamma\subset\cZ(G)$ (keeping the system fixed). Let us
  thus consider a central subgroup $\Gamma$ which is contained in
  $\Gamma_A$ such that $\Gamma\subset\Gamma_A\subset\cZ(G)$. In what
  follows we wish to argue that this inclusion of subgroups gives rise
  to a natural inclusion of topological phases. For the two symmetries
  $G_A$ and $G_\Gamma$, the topological phases are described by
\begin{align}\nonumber
  H^2\bigl(G_A,U(1)\bigr)
  \ &\cong\ \Hom\bigl(\Gamma_A,U(1)\bigr)
  \quad\text{ and }\\
  H^2\bigl(G_\Gamma,U(1)\bigr)
  \ &\cong\ \Hom\bigl(\Gamma,U(1)\bigr)\ \ .
\end{align}
  We expect that $G_A$ provides a finer resolution of topological
  phases than $G_\Gamma$. In other words, from the perspective of
  $G_\Gamma$ some of the original topological phases cannot be
  distinguished and need to be identified. It turns out that this
  identification is done via the abelian group $\Gamma_A/\Gamma$ which
  measures to which extent $\Gamma_A$ is larger than $\Gamma$. This
  suggests a relation of the form $H^2\bigl(G_\Gamma,U(1)\bigr)\cong
  H^2\bigl(G_A,U(1)\bigr)/(\Gamma_A/\Gamma)$ and indeed a simple
  calculation yields
\begin{align}\nonumber
  \label{eq:EmbeddingChain}
  \Hom\bigl(\Gamma,U(1)\bigr)
  \ &\cong\ \Hom\bigl(\Gamma_A/(\Gamma_A/\Gamma),U(1)\bigr)\\
  \ &\cong\ \Hom\bigl(\Gamma_A,U(1)\bigr)/(\Gamma_A/\Gamma)\ \ .
\end{align}
  By considering embedding chains of central subgroups, the previous
  procedure yields a whole hierarchy of topological phases.

  In the previous example it was straightforward to change the
  perspective from $G_A$ to $G_\Gamma$ with $\Gamma\subset\Gamma_A$
  and then back from $G_\Gamma$ to $G_A$. In many situations, however,
  it is even possible to change the perspective from $G_A$ to a
  smaller group $G_\Gamma$ right away. In this case the latter is
  obtained from a central subgroup $\Gamma$ satisfying
  $\Gamma_A\subset\Gamma\subset\cZ(G)$. For instance, a fixed system
  with symmetry $G$ can (under certain circumstances) be interpreted
  as a system with symmetry $PG$ (or any of the other groups
  $G_\Gamma$). This requires no modification of the physical system
  but rather a reinterpretation of its underlying Hilbert space by
  means of a blocking procedure in which several sites are combined
  into one. Under blocking, certain tensor products of
  $G_A$-representations indeed lift to a representation of
  $G_\Gamma$ since the individual projective classes (with respect to
  $\Gamma$) add up and might eventually give $[0]\in
  H^2\bigl(G_\Gamma,U(1)\bigr)$.

  For the sake of concreteness we explain the idea in a simple
  example. Most antiferromagnetic spin chains are modeled using a
  chain of on-site Hilbert spaces $\cH_k$ which are alternating
  between a representation space $\cH$ and its dual $\cH^\ast$, both
  having a well defined congruence class with respect to the action of
  $\cZ(G)$. Let us assume that the actual symmetry group is $G_A$,
  with a specific central subgroup $\Gamma_A\subset\cZ(G)$. In this
  situation, we can combine two neighboring sites $\cH$ and $\cH^\ast$
  into a single site $\cH_{\text{block}}=\cH\otimes\cH^\ast$ which
  resides in the trivial class
  $[\cH_{\text{block}}]\equiv[\cH]+[\cH^\ast]\equiv[0]$ with respect
  to $PG$. Blocking thus allows to move within the hierarchy of
  topological phases. It might happen, e.g.\ in spin ladders, that the
  Hilbert space $\cH$ decomposes into several irreducible
  representations of $G$ which belong to distinct congruence
  classes. In this situation, blocking does not give rise to a
  symmetry $PG$. Examples for hierarchies of topological phases are
  presented below in Section \ref{sc:Example}.
  
  Parts of our discussion might look very academic at first
  sight. However, there are also direct physical implications. Imagine
  two spin chains with actual symmetry groups $G_A$ and $G_B$. If we
  couple the two chains, thus building a spin ladder, the actual
  symmetry group of the complete system will be determined by the
  intersection $\Gamma_{A\cup
    B}=\Gamma_A\cap\Gamma_B\subset\cZ(G)$. In the case of $SU(2)$ spin
  ladders involving a mixture of integer and half-integer spin
  representations the intersection is trivial, thus confirming the
  observation of Ref.\ \onlinecite{Anfuso:PhysRevB.75.144420} that
  edge modes are not topologically protected.

\subsection{\label{sc:Haldane}A generalization of Haldane's
  conjecture to arbitrary groups}
  
  As we will now explain, our analysis hints towards a natural
  generalization of Haldane's conjecture. In its original formulation
  for the thermodynamic limit of the antiferromagnetic $SU(2)$
  Heisenberg Hamiltonian for spin $S$ representations, it consists of
  the following two statements:
  \cite{Haldane:1983464,Haldane:PhysRevLett.50.1153} First of all,
  there is a unique ground state which is translation
  invariant. Secondly, there is a gap above the
  ground state if $S$ is integer and the chain is gapless if $S$ is
  half-integer (i.e.\ if $2S$ is odd). Manifold evidence has been
  found to support the conjecture. In particular, it is well motivated
  in the semi-classical limit where the spin $S$ is large and where
  one can derive an effective description in terms of non-linear
  $\sigma$-models with or without $\Theta$-term.
  \cite{Haldane:1983464,Haldane:PhysRevLett.50.1153} Also, the
  absence of a gap could be
  proved using the non-trivial action of the center of $SU(2)$ on
  representations with half-integer spin.
  \cite{Lieb:1961fr,Affleck:1986pq} On the other hand, a rigorous
  mathematical proof of the existence of a mass gap for integer spins
  still seems to be open. The
  invention of the AKLT chain (in which a mass gap can be proven
  \cite{Affleck:1987cy}) was an attempt to cure this unsatisfactory
  situation. In any case, the relevance of the center of $SU(2)$ and
  of its action on specific representations already indicates a close
  relation to our present work.
 
\begin{figure}
\includegraphics[]{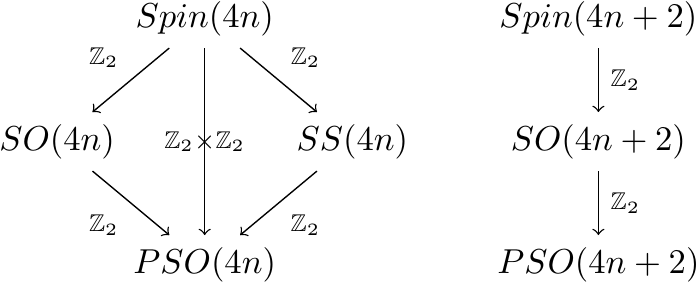}
  \caption{\label{fig:HierarchySpin} The hierarchy of
    topological phases in $Spin(2n)$ spin chains.}
\end{figure}

  Rather recently, the existence of Haldane gaps was revisited
  for different types of $SU(N)$-invariant spin chains
  \cite{Greiter:PhysRevB.75.184441,Rachel2010:PhysRevB.80.180420} (see
  Ref.\ \onlinecite{Affleck:1985wb} for some older work).
  In particular, the authors of Ref.\ 
  \onlinecite{Greiter:PhysRevB.75.184441,Rachel2010:PhysRevB.80.180420}
  claimed that $SU(N)$ chains with two-site
  interactions possess a Haldane-type gap due to spinon confinement if
  the physical sites are described by an irreducible representation
  $\lambda$ whose Young tableau possesses a number $|\lambda|$
  of boxes which can be divided by $N$. In view of our discussion in
  Section \ref{sc:SUCase} this just corresponds to the statement that
  $[\lambda]\equiv[0]$, i.e.\ the representation of $SU(N)$ needs to
  descend to a representation of $PSU(N)$. With $PSU(N)$ playing the
  same role as $SO(3)$, this suggests an obvious generalization of
  Haldane's original conjecture to an arbitrary simply-connected
  symmetry group $G$: The center $\cZ(G)$ should act trivially,
  $[\lambda]\equiv0$, in order to find a Haldane phase.
  
  However, the authors of Ref.\  
  \onlinecite{Greiter:PhysRevB.75.184441,Rachel2010:PhysRevB.80.180420}
  noted something even more interesting: A confinement similar to the
  one above can also be observed whenever $|\lambda|$ and $N$ have a
  non-trivial common divisor different than $N$. With an important
  difference to the previous case: The ground state is degenerate now
  and the interaction needs to
  encompass $N/q+1$ sites where $q=\gcd(|\lambda|,N)$. Our discussion
  of the hierarchy of topological phases immediately exhibits: Under
  the conditions specified, the representation $\lambda$ is a linear
  representation of the group $SU(N)/\Integer_q$. Since the second
  cohomology of this group is isomorphic to $\Integer_q$, this still
  gives potential edge modes the chance to transform in a non-trivial
  projective representation, thus providing a
  topological argument for the presence of a Haldane gap. Proving the
  absence of a mass gap in systems where $|\lambda|$ and $N$ do not
  have common divisors appears to be a more challenging endeavor (see,
  however, Ref.\ \onlinecite{Affleck:1986pq} for two-site interactions).
 
  An extrapolation of our previous arguments suggests that spinon
  confinement (for a suitable interaction range) exists if and only if
  the physical system allows for a non-trivial way of enhancing its
  symmetry at (virtual) edges. Equivalently, the physical Hilbert
  spaces $\cH_k$ have to belong to the trivial congruence class
  $[\cH_K]_\Gamma\equiv[0]$ with respect to at least one non-trivial
  central subgroup $\Gamma\subset\cZ(G)$ such that the relevant
  symmetry of the system is $G_\Gamma$, a proper quotient of $G$.
  For matrix product states, the existence or absence of a mass gap
  (with respect to a specific model Hamiltonian) is intimately related
  to the possibility of realizing it in an ``injective'' way.
  \cite{Perez-Garcia:2007:MPS:2011832.2011833,Sanz:PhysRevA.79.042308}
  Most likely, a suitable adaption of these arguments provides the
  route for a proof of our statement.

  A non-trivial test of our conjecture should be possible along the
  lines of Ref.\ 
  \onlinecite{Greiter:PhysRevB.75.184441,Rachel2010:PhysRevB.80.180420} for
  the groups $Spin(4n)$, see Section \ref{sc:SOCase}. In this case the
  center is given by $\Integer_2\times\Integer_2$ and it admits three
  inequivalent embeddings
  $\Integer_2\subset\Integer_2\times\Integer_2$, either into the left
  or right factor or diagonally. It turns out that among the three
  quotients $Spin(4n)/\Integer_2$ two are isomorphic, leading to the
  so-called semispinor group $SS(4n)$, while the remaining one is
  isomorphic to $SO(4n)$ (but not isomorphic to $SS(4n)$ as long as
  $n\neq2$). \cite{Mimura:MR1122592} The resulting hierarchy of
  quotients is displayed in Figure \ref{fig:HierarchySpin}. One can
  thus imagine to build spin chains based on linear representations of
  $SO(4n)$ or $SS(4n)$ which are only projective representations of
  $PSO(4n)$. It is likely that some of these chains would enjoy
  topological protection, resulting in non-trivial edge modes
  transforming in a projective representation of $SO(4n)$ or $SS(4n)$,
  respectively. A priori, it is not clear whether gapped spin
  chains of this type can be realized with two-site
  interactions. Block renormalization and the experience with $SU(N)$,
  however, suggests that such spin chains should exist if interactions
  across several sites are permitted. Similar remarks apply to
  $Spin(4n+2)$ which has a non-trivial central subgroup
  $\Integer_2\subset\Integer_4$.

\begin{figure}
\includegraphics[]{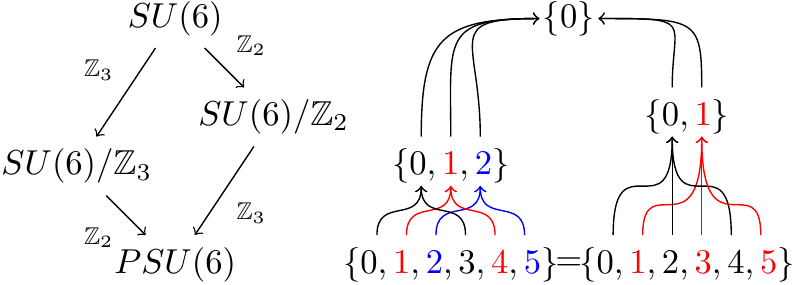}
  \caption{\label{fig:Hierarchy}(Color online) The hierarchy of
    topological phases in $SU(6)$ spin chains.}
\end{figure}

\subsection[Two illustrative examples: $SU(6)$ and $SU(12)$ spin
chains]{\label{sc:Example}Two illustrative examples: \texorpdfstring{\textit{SU}(6)}{SU(6)}
  and \texorpdfstring{\textit{SU}(12)}{SU(12)} spin chains}

  In this section we wish to focus on spin chains with
  $SU(6)$-symmetry. This example nicely illustrates the technical
  aspects and the physical implications of our work. The group $SU(6)$
  has center $\Integer_6$. We have three different choices for
  non-trivial subgroups $\Gamma$. Either we choose $\Integer_2$,
  $\Integer_3$ or the full group $\Integer_6$ itself. Depending on the
  choice of physical Hilbert spaces $\cH_k$, one then ends up with one
  of four symmetry groups: $SU(6)$, $PSU(6)$, $SU(6)/\Integer_2$ or
  $SU(6)/\Integer_3$.

  The topologically richest systems are those with $PSU(6)$
  symmetry. In this case we expect six different topological phases
  which manifest themselves in the congruence class
  $[\cB]\in\Integer_6$ of (virtual) edge modes. They are labeled by
  $[\cB]\in\{[0],[1],[2],[3],[4],[5]\}$. In systems with
  $SU(6)/\Integer_3$-symmetry we still have three distinct topological
  phases, which are labeled by
  $[\cB]_{\Integer_3}\in\{[0],[1],[2]\}$. Since the center of
  $SU(6)/\Integer_3$ is isomorphic to $\Integer_2$ and the double
  quotient gives rise to
  $\bigl(SU(6)/\Integer_3\bigr)/\Integer_2=PSU(6)$, the phases of
  $PSU(6)$ can be identified with the phases of $SU(6)/\Integer_3$ up
  to the identifications $[0]\sim[3]$, $[1]\sim[4]$ and
  $[2]\sim[5]$, thus $[\cB]_{\Integer_3}\equiv[\cB]\mod3$. Conversely,
  if we have a topological phase $[\cB]_{\Integer_3}$ there is a
  chance (but no need) that it admits an interpretation as a phase of
  type $[\cB]$ or $[\cB]+[3]$ in an $PSU(6)$-chain.

\begin{figure}
\includegraphics[width=\columnwidth]{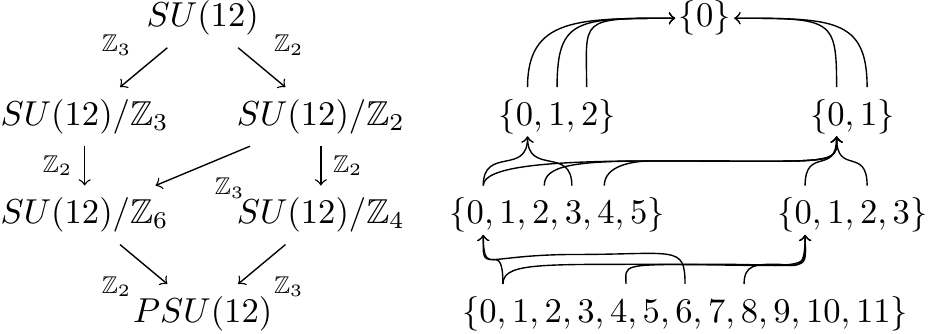}
  \caption{\label{fig:Hierarchy12} The hierarchy of
    topological phases in $SU(12)$ spin chains.}
\end{figure}

  Similarly, a system with $SU(6)/\Integer_2$ has two distinct
  topological phases labeled by
  $[\cB]_{\Integer_2}\in\{[0],[1]\}$. Now we have
  $PSU(6)=(SU(6)/\Integer_2)/\Integer_3$ and
  $[\cB]_{\Integer_2}\equiv[\cB]\mod2$. The whole
  hierarchy of topological phases for $SU(6)$ is depicted in Figure
  \ref{fig:Hierarchy}. We can easily confirm that Haldane phases
  should exist for representations with
  $[\cB]\in\{[0],[2],[3],[4]\}$ albeit they are protected by
  different symmetries. These numbers are precisely those having
  non-trivial common divisors with $6$ (the $6$ of $SU(6)$), in accord
  with the results of Ref.\  
  \onlinecite{Greiter:PhysRevB.75.184441,Rachel2010:PhysRevB.80.180420}. 
  They are represented in black color in the lower line of Figure
  \ref{fig:Hierarchy}. For higher rank groups the hierarchies becomes
  more involved, but they can be derived following the same
  principles. In Figure \ref{fig:Hierarchy12} the hierarchy for the
  group $SU(12)$ is depicted. The extra structure arises from the fact
  that $\mathbb{Z}_{12}$ has subgroups, for example $\mathbb{Z}_{3}$
  and $\mathbb{Z}_{4}$, that are not subgroups of each other.

\section{\label{sc:ColdAtoms}Application to cold atom systems}

  The final Section of our paper is devoted to the application of our
  general formalism to the study of quantum magnetism in cold atom
  systems. The continuous symmetries relevant in this context are
  $SP(4)$ (or, equivalently,
  $Spin(5)$) and $SU(N)$, with even values of $N$ up to
  $N=10$.\cite{Hung:2011PhRvB..84e4406H,2010NatPh...6..289G,Nonne:2012arXiv1210.2072N}
  In what follows we shall focus on the series $SU(N)$. We first
  outline how the Heisenberg Hamiltonian arises as a particular limit
  of a Fermi-Hubbard model. Afterwards we discuss how particular
  examples fit into our general framework.

\subsection{\label{sc:Heisenberg}The \texorpdfstring{\textit{SU(N)}}{SU(N)} Heisenberg model
  from cold atoms}

  The realization of an $SU(N)$ symmetry requires a large number of
  degenerate energy levels. As was emphasized in Ref.\
  \onlinecite{2010NatPh...6..289G}, the latter arise naturally
  in earth-alkaline atoms. Since the nuclear spin $I$ reaches values
  up to $I=9/2$ (for $^{87}\text{Sr}$), one can easily achieve
  degeneracies up to $2I+1=10$. The resulting states can be identified
  with the $N$-dimensional fundamental representation of $SU(N)$, with
  $N=2I+1$. Earth-alkaline systems exhibit an almost perfect
  decoupling of nuclear and electronic spin degrees of freedom. In
  practice, this means that the degeneracy is not lifted by
  interactions. For this reason, the $SU(N)$ symmetry is still
  reflected in the Hamiltonian describing the dynamics of the atoms in
  an optical lattice. Effectively, one thus arrives at an $SU(N)$
  symmetric Fermi-Hubbard model. Similar to
  the familiar case of the Mott insulator phase, there exists a
  certain parameter range where the model can be approximated in terms
  of an $SU(N)$ anti-ferromagnetic Heisenberg spin
  chain.\cite{2010NatPh...6..289G,Nonne:2012arXiv1210.2072N}

\subsection{\label{sc:Application}Realization of topologically
  non-trivial phases}

  For the physics of the system, it is essential to know the $SU(N)$
  representation on which the spin operators act. This representation
  is determined by the occupation number per
  site.\cite{2010NatPh...6..289G,Nonne:2012arXiv1210.2072N} The
  situation that will be of interest for us is the two-orbital case at half-filling, i.e., with $N$ atoms per site.
  As was argued in Ref.\ \onlinecite{Nonne:2012arXiv1210.2072N}, the
  relevant $SU(N)$ representation $\lambda$ is then specified by a
  Young tableau with two columns and $N/2$ rows. Using the
  general formula \eqref{eq:SUClassYoung} we find that
  $[\lambda]\equiv[0]$. Accordingly, $\lambda$ cannot only be
  interpreted as a representation of $SU(N)$ but it also descends to
  the quotient group $PSU(N)=SU(N)/\Integer_N$. It is thus natural to
  ask which of the $N$ possible topological phases is actually
  realized by the cold atom system.

  The authors of Ref.~\onlinecite{Nonne:2012arXiv1210.2072N}
  argued that the system realizes a topologically
  non-trivial phase. This claim was supported by the existence of
  AKLT-type Hamiltonians which act on the same physical Hilbert space
  and which are utilizing an auxiliary representation $\cB$ which is
  described by
  a Young tableau with $N/2$ rows in a single column. With our formula
  \eqref{eq:SUClassYoung} we easily verify that $[\cB]\equiv[N/2]$,
  i.e.\ the AKLT-type system indeed corresponds to a non-trivial topological
  phase. Since the AKLT-type Hamiltonian for $N=4$ provides a close
  approximation to the Heisenberg Hamiltonian, the same non-trivial
  topology was conjectured for the cold atom system in the relevant
  range of parameters.\cite{Nonne:2012arXiv1210.2072N}

  At this moment of time, it is still an open question whether the
  Heisenberg Hamiltonian and the AKLT-type Hamiltonian really belong
  to the same topological phase. On the other hand, it is known that
  the topological phase can be extracted unambiguously from a suitable
  string order parameter.\cite{Duivenvoorden:2012} Our current work
  thus provides an important step towards settling this crucial
  issue. Moreover, it suggests the existence of other topological
  phases of $PSU(N)$ spin chains which might be realizable in cold
  atom systems. A more detailed discussion of these aspects will be
  reported elsewhere.

\section{\label{sc:Conclusions}Conclusions}

  In our paper we revisited the classification of topological phases
  in gapped spin chains with continuous symmetry group. We identified
  and evaluated the relevant cohomology groups
  $H^2\bigl(G_\Gamma,U(1)\bigr)$ and showed that
  they are isomorphic to the central subgroup $\Gamma\subset\cZ(G)$
  defining $G_\Gamma=G/\Gamma$ as a quotient of its simply-connected
  cover $G$. For a number of symmetries, among them $PSU(N)$ and
  $PSO(2N)$, we found more than one topologically non-trivial
  phase. In particular, we wish to emphasize the remarkable fact that
  for $PSU(N)$ the number of topological phases is $N$ and hence
  increases with the rank of the symmetry group. For the projective
  groups $PG=G/\cZ(G)$ a complete summary of our classification result
  can be read off from Table \ref{tb:CongruenceClasses}. The
  cohomology groups $H^2\bigl(G_\Gamma,U(1)\bigr)$ exhibit
  mathematical relations when considered for different choices of the
  subgroup $\Gamma\subset\cZ(G)$. These dependencies lead to a natural
  hierarchy of topological phases. In Section \ref{sc:Physics} we
  managed to explain this hierarchy from a physical perspective by
  considering blocking procedures and the combination of spin chains
  into spin ladders.

  Our classification of topological phases -- and the distinguished
  role played by the central subgroups $\Gamma\subset\cZ(G)$ -- led us
  to propose a natural generalization of Haldane's conjecture
  \cite{Haldane:1983464,Haldane:PhysRevLett.50.1153} to arbitrary
  symmetry groups, see Section \ref{sc:Haldane}. In our more general
  setup, the original distinction between half-integer and integer
  spin $S$ of $SU(2)$ is replaced by whether a representation
  $\lambda$ is a linear representation of {\em any} of the groups
  $G_\Gamma$ (i.e.\ $[\lambda]_\Gamma\equiv[0]$) where
  $\Gamma\subset\cZ(G)$ is a non-trivial central subgroup of $G$. Our
  proposal is in complete accord with a recent analysis of Haldane
  phases in $SU(N)$ spin chains by Greiter and Rachel.
  \cite{Greiter:PhysRevB.75.184441} We believe that their analysis
  can be carried over to groups of type $Spin(2N)$, thus providing a
  non-trivial check of our conjecture.

  Let us briefly discuss the implications of our results for the study
  of concrete physical systems, possibly from a numerical point of
  view. In our opinion, it cannot be overemphasized that in
  many spin chains there are {\em more than two distinct} topological
  phases. While it is a relatively simple task to distinguish between
  a topologically trivial and a non-trivial phase, e.g.\ using a
  suitable string order parameter \cite{DenNijs:PhysRevB.40.4709} (for
  a general discussion see Ref.\
  \onlinecite{PerezGarcia:PhysRevLett.100.167202}), the definition of
  a quantity which can be calculated efficiently and which can
  discriminate between all different topologically non-trivial phases
  is still an open problem. Significant progress with regard to such
  order parameters has recently been made in Ref.\
  \onlinecite{Haegeman:1201.4174v1,Pollmann:2012PhRvB..86l5441P}. However,
  both papers focused on discrete symmetries and an application of
  similar ideas to the cases at hand remains to be worked out. In a
  companion paper \cite{Duivenvoorden:2012} we will fill this gap and
  provide an explicit expression for a string order parameter for
  $SU(N)$ spin chains which can easily be evaluated once the ground
  state is known. It will be proven that our order parameter is
  sensitive to the projective class describing the topological phase
  and that it allows to discriminate all $N$ distinct phases of
  $PSU(N)$ spin chains. The string order parameter may therefore be
  used to verify the claim of
  Ref.~\onlinecite{Nonne:2012arXiv1210.2072N} that non-trivial
  topological phases of $PSU(N)$ spin chains can be simulated in cold
  atom systems, see also Section \ref{sc:ColdAtoms}.

  Our analysis calls for extensions in several directions. First of
  all, our classification was concerned with continuous on-site
  symmetries only. Taking into account additional discrete symmetries
  such as translation symmetry, time-reversal symmetry or inversion
  symmetry will modify the classification.
  \cite{Chen:PhysRevB.83.035107,Chen:PhysRevB.84.235128} In order to
  gain some intuition for the underlying reasons, let us briefly
  discuss the effects of imposing either time-reversal or inversion
  symmetry (or both), in addition to the on-site symmetry
  $G$. According to Ref.\ \onlinecite{Chen:PhysRevB.84.235128}, apart
  from the cohomology groups $H^2\bigl(G,U(1)\bigr)$ another important
  ingredient is the space of one-dimensional representations of
  $G$. For simple Lie groups $G$, the only one-dimensional
  representation is the trivial representation. Hence this data does
  not give rise to additional topological phases in our situation.

  On the other hand, it was observed that the projective class
  $[\lambda]$ describing the boundary modes has to satisfy
  $2[\lambda]\equiv0$ in the presence of either inversion or
  time-reversal symmetry. This leads to a possible reduction in the
  number of topological phases. Actually, the constraint
  $2[\lambda]\equiv0$ can be understood quite easily from the matrix
  product state construction reviewed in Section \ref{sc:MPS}. It is
  obvious for instance that inversion symmetry requires the auxiliary
  spaces to be self-conjugate, $\lambda=\lambda^+$, since they are
  exchanged under inversion. In view of the general relation
  $[\lambda^+]=-[\lambda]$, this condition immediately implies
  $2[\lambda]\equiv0$. Similar remarks apply to time-reversal.

  As we have just seen, enforcing the presence of additional
  symmetries may drastically reduce the number of topological phases
  which can exist in spin chains with continuous symmetry. In
  particular, for $PSU(N)$ there are no non-trivial inversion
  symmetric topological phases if $N$ is odd. Indeed, the construction
  of the two non-trivial topological phases in an $PSU(3)$ spin chain
  that was presented in Ref.\ \onlinecite{Duivenvoorden:2012}
  explicitly required to break inversion symmetry. On the other hand,
  there is precisely one topologically non-trivial inversion symmetric
  phase if $N$ is even. An explicit realization of this phase has been
  constructed in Ref.\ \onlinecite{Nonne:2012arXiv1210.2072N}.
  Using the results of Ref.\ \onlinecite{Chen:PhysRevB.83.035107} and
  our own classification it is a straightforward exercise to work out
  all topological phases which are protected by a combination of
  continuous on-site symmetries $G$ and/or time-reversal or inversion
  symmetry.

  Another interesting open point concerns the interplay of continuous
  symmetries with discrete internal symmetries, arising e.g.\ in spin
  ladders. The presence of these additional symmetries will lead
  to adjustments (see e.g.\ Ref.\
  \onlinecite{Liu:2012arXiv1204.5162L}) which require a separate
  analysis, depending on the precise type of model under
  consideration. We believe that the results presented here will be
  helpful in accomplishing this task.

  It seems feasible to generalize our results to supersymmetric and
  $q$-deformed spin chains. We hope to report on this in the near
  future. On the other hand, an extension to non-compact groups
  appears to be more challenging from a technical point of view. While
  the mathematical part of the story -- the topology of non-compact
  groups and the division of representations into congruence classes
  -- seems to be well understood, the complications arise on the
  physical side. In particular, it is evident that non-compact groups
  come hand in hand with infinite dimensional representations,
  together with all their functional analytic intricacies. For
  example, it is not clear to us at the moment whether in the infinite
  dimensional setup symmetry preserving matrix product states can be
  constructed which admit a parent Hamiltonian describing a gapped
  phase.

\begin{acknowledgments}
  We gratefully acknowledge useful discussions with Harm Michalski,
  Achim Rosch and Daniel Wieczorek.
  The work of Kasper Duivenvoorden is funded by the German Research
  Foundation (DFG) through the SFB$|$TR\,12 ``Symmetries and
  Universality in Mesoscopic Systems'' and the ``Bonn-Cologne Graduate
  School of Physics and Astronomy'' (BCGS). The work of Thomas Quella
  is funded by the DFG through Martin Zirnbauer's Leibniz Prize, DFG
  grant no.\ ZI 513/2-1.
\end{acknowledgments}



\end{document}